\documentclass[10pt]{article}
\usepackage{amssymb,amsmath}
\usepackage{graphicx,xcolor}
\usepackage{floatrow,caption}
\usepackage{subfig}
\usepackage{mathrsfs}
\usepackage{hyperref}
\hypersetup{colorlinks=true,allcolors=[rgb]{0,0,0.6}}
\newtheorem{thm}{Theorem}[section]
\newtheorem{lem}[thm]{Lemma}
\newtheorem{prop}[thm]{Proposition}
\def\proof{{\bf Proof. }}
\def\bea{\begin{eqnarray}}
\def\eea{\end{eqnarray}}
\def\bean{\begin{eqnarray*}}
\def\eean{\end{eqnarray*}}
\def\Z{{\mathscr{Z}}}
\def\P{{\mathscr{P}}}
\def\L{{\mathscr{L}}}
\def\S{{\mathcal{S}}}

\newcommand{\field}[1]{\mathbb{#1}}
\newcommand{\nz}{\field{N}}

\def\ra{{\rangle}}
\def\la{{\langle}}
\def\fin{{$\hfill\square$}}
\def\hbarr{{\varepsilon}}

\def\Tr{{\rm{Tr}}}

\begin{document}
\title{Mean field approximation of many-body quantum dynamics for Bosons in a discrete numerical model}
\author{B.~Pawilowski\thanks{IRMAR, Universit{\'e} de Rennes I, UMR-CNRS 6625, Campus de Beaulieu, 35042 Rennes Cedex, France}
\thanks{Fak. Mathematik, Univ. Wien, Oskar-Morgenstern-Platz 1, 1090 Wien, Austria}}
\date{\today}

\maketitle
\begin{abstract}
 The mean field approximation is  numerically validated in the bosonic
 case by considering the time evolution of quantum states and their associated
reduced density matrices by many-body Schr\"odinger dynamics. The
model phase-space is finite-dimensional. The results are illustrated with numerical simulations of the
evolution of quantum states according to the time, the number of
the particles, and the dimension of the phase-space.
\end{abstract}


\section{Introduction}
\label{sec.Intro}
The mean field approximation is known to be a good way to approximate the
many-body Schr\"odinger dynamics when the number of particles is
large enough (see \cite{AmBr,AmNi4,BGM,BEGMY,ChPa,ElSc,ErYa,ESY1,ESY2,FGS,FKP,FKS,GMP,KlMa,LSSY,Pi,GiVe1,GiVe2,Hep,Spo}).
\\
It consists in looking for the solutions to the non-linear Schr\"odinger equation
for one particle called the Hartree equation.
We are interested in the density matrix associated with the
wave function, this matrix satisfies the quantum Liouville equation
dual to the Von Neumann equation.
The partial trace operators of this matrix, called the
reduced density matrices, satisfy a hierarchy of
equations.
For instance, by considering the case where the initial state for the
Schr\"odinger equation is a Hartree ansatz(a product state)
which is suitable for a bosons condensate, the
limit, when the number of particles $N$ goes to the infinity of these matrices converge
in trace norm to the product of the density matrix associated
with the solution to the Hartree equation.
And this asymptotic density matrix satisfies the time dependent
Hartree equation \cite{BaMa}.
\\
When the particles are bosons, the
suitable space for the bosons is the symmetric Fock space on
the phase-space. Moreover for the sake of numerical computations, a finite-dimensional phase-space will be used instead of an usual
phase-space of type $L^2(\mathbb R^d)$. So here the phase-space
will be $\mathcal Z = \ell^2(\{0,\cdots,K\}) \simeq \mathbb C^K$ where $K$ is a given
integer representing the number of sites. Each particle can live in
one of the $K$ sites.
\\
For the numerical implementation, an explicit basis of the $N$-fold
sector of the Fock bosonic spaces is specified. This basis allows the
numerical computation of the full $N$-body quantum problem for $N$
large enough to validate various mean field regimes, in spite of a
rapidly increasing complexity.
\\
The resolution of the $N$-particles Schr\"odinger equation will
rely on a splitting method, one part for the free Hamiltonian
and the other one for the two particles interaction term.
\\
For the simulations, the considered real bounded potential associated
with the interaction term will be $V$ defined on $\mathbb Z / K\mathbb
Z$ by $V(i)=\frac 1{|i|}$ if $i\neq 0$ and $V(0)=0$.
\\
According to previous results related to the propagation of the Wigner
measures \cite{AmNi1,AmNi2,AmNi3,AmNi4} knowing the Wigner measure at time $t=0$
determines the Wigner measure at time $t$ and all asymptotic reduced density matrices. 
For many examples, like Hermite states, twin Fock states or states
studied in quantum information theory (see \cite{AFP}) their Wigner
measure as well as the order of convergence of reduced density
matrices is known explicitly. The evolution of the Wigner measure,
and consequently of the asymptotic density matrices, is evaluated
after integrating numerically the mean field non linear Hartree
time-dependent equation. In order to preserve numerically quadratic quantities like the symplectic form, the latter is solved with a symplectic $4^{th}$ order Runge-Kutta method (\cite{HLW}).
\\
To estimate numerically the error of convergence of the reduced
density matrices in the mean field limit, a discretization
of a time interval $[0,t_{max}]$ is considered, in the examples
$t_{max}=1$ is chosen, then the quantity
 $\max_{t\in[0,t_{max}]}\left\|\gamma^{(p)}_{N}(t)-\gamma_{\infty}^{(p)}(t)\right\|_1$
 is observed. Here $\gamma^{(p)}_{N}(t)$ denotes the time-evolved $p$ particles reduced density
matrix for $N$ bosons, while $\gamma_{\infty}^{(p)}(t)$ is its theoretical
limit when $N$ goes to the infinity, and $\|.\|_1$ denoting the trace norm.
\\
For the evaluation of the order of convergence, the logarithm of the
previous error estimate is drawn according to $log(N)$. This gives a straight line
whose the slope is the order of convergence in $1/N$.
\\
These numerical results agree very well and illustrate the theoretical
analysis carried out in \cite{AFP}.
\\
By increasing $K$, we wish to approach a continuous model. The complexity of the computations increase in the same time that
$K$ and $N$ increase because of the dimension of the $N$-particles
bosons Fock space on $\mathbb C^K$ which is a binomial coefficient.

\section{Framework}
\label{sec.framework}

 The bosons Fock space on an Hilbert space $\mathcal Z$ is defined as
 $\Gamma_{s}(\mathcal Z)=\bigoplus_{n \geq 0}\bigvee^{n}\mathcal Z$ where $\bigvee^{n}\mathcal Z$ is the symmetric
n-fold Hilbertian tensor product of $\mathcal Z$ which is the range of
the projection defined on the Hilbert tensor product $\mathcal
Z^{\otimes n}$, by:

$$
S_{n}(\xi_{1}\otimes
\xi_{2} \otimes ... \otimes \xi_{n})=\frac{1}{n!}\sum_{\sigma \in
  \Sigma_{n}}\xi_{\sigma(1)} \otimes \xi_{\sigma(2)} \otimes ... \otimes
\xi_{\sigma(n)},
$$
where $\xi_i$ is in $\mathcal Z$ for each $i$ in $[1,n]$ and $\Sigma_n$ is the set of the permutations of $n$ elements.\\

 For $z$ in $\mathcal Z$ and $\varepsilon$ positive, the
 $\varepsilon$-scaled annihilation
 and creation operators are defined for all $\Phi$ in $\mathcal Z$ and
 $n$ in $\mathbb N$ by:
\begin{align*}
a(z)\Phi^{\otimes n}&= 
\sqrt{\varepsilon n} \langle z| \Phi \rangle\Phi^{\otimes n-1},\\
 a^{*}(z)\Phi^{\otimes n}&=\sqrt{\varepsilon (n+1)} S_{n+1} (
|z\rangle \otimes\Phi^{\otimes n}).
\end{align*} 
These operators are then extended by linearity and density to $\bigvee^{n}\mathcal{Z}$.\\

These operators satisfy the canonical commutation relations (CCR):
\begin{align} \label{CCR}
[a(z_1),a^{*}(z_2)]&=\varepsilon \langle z_1,z_2\rangle Id\;, &\\
[a(z_1),a(z_2)]&=0\;, &\\
[a^{*}(z_1),a^{*}(z_2)]&=0\;. 
\end{align}

The  second quantization of an operator $A \in \mathcal L (\mathcal Z)$
or a self-adjoint operator $(A,D(A))$ in $\mathcal Z$ is defined by:
\begin{equation*}
d\Gamma(A)_{|\vee^{n,alg}D(A)}=\varepsilon  \sum_{i=1}^{n}Id^{\otimes
  i-1} \otimes A \otimes Id^{\otimes n-i}\;.
\end{equation*}
The second quantization of $Id_{\mathcal Z}$ is the number operator:
\begin{equation*}
\mathbf{N}_{|\vee^{n}\mathcal Z}=\varepsilon n Id_{\vee^{n}\mathcal Z}\;.
\end{equation*}

\subsection{Orthogonal basis of the $N$-fold sector}
\label{sec.basis}

Use the following notations: $\mathbb Z_{K}=\mathbb Z/K\mathbb Z\;.$\\
For $\alpha=(\alpha_1,\cdots,\alpha_K)$ in $\mathbb{N}^K$, the length of $\alpha$ is written $|\alpha|=\alpha_1+\cdots+\alpha_K$ and the factorial of $\alpha$ $\alpha!=\alpha_1!\cdots\alpha_K!\;.$
\\
Let $(e_1,\cdots,e_K)$ be an orthonormal basis of $\mathbb{C}^K$.\\
Set $\mathcal Z=\mathbb C^K$. Then an orthonormal basis of
$\bigvee^{N}\mathcal Z$ can be built from this basis which is labelled
by the multi-indices $\alpha$ in $\mathbb N^K$ such that $|\alpha|=N$.\\

With the creation operators, an orthonormal basis can be written as:\\

\begin{align*}
 e_{\alpha}:=\frac{
   a^*(e)^{\alpha}}{\sqrt{\varepsilon^{|\alpha|}\alpha!}}|\Omega\rangle
 := \frac{1}{\sqrt{\varepsilon^{|\alpha|}\alpha!}} a^*(e_1)^{\alpha_1}
 \cdots {a^*(e_K)}^{\alpha_K}|\Omega\rangle\,,
\end{align*}
where $|\Omega\rangle=(1,0,0,0, \ldots)$ is the vacuum of the Fock space.\\

Then the dimension of $\bigvee^{N}\mathcal Z$ is
$card(\{\alpha \in \mathbb{N}^K/|\alpha|=N\})=\dbinom{N+K-1}{K-1}\;.$\\
And $card(\{\alpha \in \mathbb{N}^K / |\alpha|\leq
N\})={N+K \choose K}$ is the dimension of $\bigoplus_{n=0}^{N}
\bigvee^{n}\mathcal Z\;.$

\subsection{Hamiltonian}
\label{sec.Ham}
As a binomial number, the dimension of the $N$ particles bosonic
sector increases rapidly but not too much, as $N$ increases (see Table
\ref{dimensions} for numerical values).
 The complexity has to be handled carefully if we want to approach the mean
field limit by taking $N$ large or the continuous model by taking $K$ large.
\\
Define $\Delta_{K}$ the discrete Laplacian operator on $\mathbb{C}^K$ by:
$$
\forall z \in \mathbb{C}^K \;\;, \quad \forall i \in \mathbb
Z/K\mathbb Z,\quad (\Delta_{K}z)_i=z_{i+1}+z_{i-1}\;.
$$

And let $H_0=d\Gamma(-\Delta_{K})$ be the free Hamiltonian.\\

 The interaction term denoted by $\mathcal{V}$ equals:
$$
\mathcal{V}=\frac{1}{2}\sum_{(i,j)\in\mathbb{Z}_{K}^{2}}V_{ij}a^*(e_i)a^*(e_j)a(e_i)a(e_j)\;,
$$
where $V_{ij}=V_{ji}=V(i-j)$.\\
In this framework changing the value of $V(0)$ add an irrelevant phase 
factor in the time evolved wave function. In the sequel $V(0)=0$ is assumed.\\
Or as a Wick quantized operator \eqref{eq.wick}:
$$
\mathcal{V}=\big \langle z^{\otimes
  2},\big(\frac{1}{2}\sum_{(i,j)\in\mathbb{Z}_{K}^{2}}V_{ij}|e_i\otimes
e_j\rangle\langle e_i \otimes e_j|\big)z^{\otimes 2 }\big\rangle ^{Wick}\;.
$$

The considered linear Schr\"odinger equation is:
\begin{align}\label{eq.Schr}
\mathrm i\varepsilon \partial_{t}\Psi=H_{\varepsilon}\Psi \;,
\end{align}
where the complete Hamiltonian is defined on the bosonic Fock space by :
$$H_{\varepsilon}=d\Gamma(-\Delta_{K}) + \mathcal{V} \;.$$

\section{Finite dimensional mean field equation}

\subsection{Energy of the Hamiltonian}
The energy of the Hamiltonian corresponds to the symbol of the
complete Hamiltonian:
$$H(z,\bar{z})=\langle z , -\Delta_{K} z \rangle+\frac{1}{2}\sum_{i\neq j}V_{ij}|z_i|^2|z_j|^2 \;,$$
while recalling our convention $V_{ii}=V(0)=0$ for all $i$ in $\mathbb{Z}_{K}$.\\

\subsection{Hartree equation}
The mean field equation in $\mathcal Z$ is written as:
$$\mathrm i \partial_{t}z=\partial_{\bar z}H(z,\bar{z}) \;.$$
For each component in $\mathbb C^K$ we obtain:
\begin{align*}
\mathrm i \partial_{t}z_i&=\partial_{\bar z_i}H=\partial_{\bar z_i}(\sum_{i'}|z_{i'}-z_{i'-1}|^2-2|z_{i}|^{2}+\frac{1}{2}\sum_{i' \neq
  j}V_{i'j}|z_{i'}|^2|z_j|^2)\\
&=\partial_{\bar z_i}(\sum_{i'}(\bar{z_{i'}}-\bar{z_{i'-1}})(z_{i'}-z_{i'-1})-2 \bar{z_i}z_i+\frac{1}{2}\sum_{i'
  \neq j} V_{i'j}\bar{z_{i'}}z_{i'}\bar{z_j}z_j)\\
&=z_i-z_{i-1}-(z_{i+1}-z_i) -2z_i+\sum_{j \neq i}V_{ij}z_i|z_j|^2\\
&=-(z_{i+1}+z_{i-1})+(\sum_{j \neq i}V_{ij}|z_j|^2)z_i
\end{align*}

$$\dot z_i=-\mathrm i[-(\Delta_{K}z)_i+(\sum_{j \neq i}V_{ij}|z_j|^2)z_i]$$

By writing $z=q+\mathrm i p$ where $q$ and $p$ belong to $\mathbb
R^K$, it becomes:
\begin{eqnarray*}
\dot q_i &=&-(p_{i+1}+p_{i-1})+\sum_{j\ne i} V_{ij} (q_j^2+p_j^2)p_i\\
\dot p_i &=& -\Big( -(q_{i+1}+q_{i-1})+\sum_{j\ne i} V_{ij} (q_j^2+p_j^2)q_i\Big)
\end{eqnarray*}
\\
A $4^{th}$ or $6^{th}$ order Gauss RK method is used by using the coefficients
given in \cite{HLW} to solve the Hartree equation.\\ 
A symplectic method is used to preserve the quadratic part of the
energy, the symplectic form and the phase space volume. 

\subsection{Wigner measures}
\label{sec.wigner-meas}
For $f \in \mathcal Z$ the field operator is defined by $\Phi(f)=\frac{1}{\sqrt{2}}(a^{*}(f)+a(f))$ which is essentially self-adjoint on $\Gamma_{fin}(\mathcal{Z})=\bigoplus_{n\in\mathbb N}^{alg}\bigvee^{n}\mathcal{Z}$\\
The Weyl operator is defined by $W(f)=e^{i\Phi(f)}$.\\

Let $(\varrho_{\varepsilon})_{\varepsilon \in \mathcal{E}}$ be a
family of normal states on $\Gamma_{z}(\mathcal{Z})$ with 
$\mathcal{E}\subset (0,+\infty)$, $0\in\overline{\mathcal{E}}$.\\
A measure $\mu$ is a Wigner measure for this family, $\mu\in
\mathcal{M}(\varrho_{\varepsilon}, \varepsilon \in \mathcal{E})$, if
there exists $\mathcal{E}'\subset \mathcal{E}$, $0\in
\overline{\mathcal{E}'}$ such that
$$
\forall f\in \mathcal{Z}\,, \lim_{\varepsilon\in \mathcal{E}'\,,
  \varepsilon\to 0}\Tr\left[\varrho_{\varepsilon}W(\sqrt{2}\pi
  f)\right]=\int_{\mathcal{Z}}
e^{2i\pi\textrm{Re}~\langle f,z\rangle}~d\mu(z) \;,
$$
see \cite{QB}.\\
The following result valid for separable Hilbert spaces $\mathcal Z$,
apply to our finite dimensional $\mathcal Z \simeq \mathbb C^K$.

 \begin{thm}\cite{AmNi1}.
\label{th.Wigdef}
If $(\varrho_{\varepsilon})_{\varepsilon \in \mathcal{E}}$
satisfies the uniform estimate 
$\Tr[\varrho_{\varepsilon}\mathbf{N}^{\delta}] \leq{C_{\delta}}<
+\infty$ for some $\delta>0$ fixed,
$\mathcal{M}(\varrho_{\varepsilon},\varepsilon\in \mathcal{E})$  is
not empty and made of Borel probability measures ($\mathcal{Z}$ separable) such
that $\int_{\mathcal Z}|z|^{2\delta}d\mu(z)\leq C_{\delta}$.
\end{thm}

For each $p$ in $\mathbb N$, the reduced density matrix associated with a state
$\varrho_{\varepsilon}$ is a trace class operator in $\mathcal
L^1(\bigvee^p \mathcal Z)$ defined by the duality relation:
\begin{align}\label{rdm-duality}
\Tr\left[\gamma_{\varepsilon}^{p}
\tilde
b\right]=\frac{\Tr\left[\varrho_{\varepsilon}b^{Wick}\right]}{
\Tr\left[\varrho_{\varepsilon}(|z|^{2p})^{Wick}\right]}
\end{align}
where $\tilde b \in \mathcal L(\bigvee^p \mathcal Z)$.
\\
The asymptotic reduced density matrix associated with the Wigner
measure $\mu$ equals:
\begin{align}\label{rdm-limit}
\gamma_{0}^{p}=\frac{\int_{\mathcal Z}|z^{\otimes
    p}\rangle \langle z^{\otimes p}|d\mu(z)}{\int_{\mathcal
    Z}|z|^{2p}d\mu(z)}\,.
\end{align}

In finite dimension, if the family $(\varrho_{\varepsilon})_{\varepsilon\in \mathcal{E}}$
satisfies $\mathcal{M}(\varrho_{\varepsilon},\varepsilon\in
\mathcal{E})=\left\{\mu\right\}$ then the $(PI)$-condition (see \cite{AmNi1}):
$$
\forall p\in \mathbb{N}\,, \lim_{\varepsilon\in
  \mathcal{E},\varepsilon\to
  0}\Tr\left[\varrho_{\varepsilon}\mathbf{N}^{p}\right]=\int_{\mathcal{Z}}|z|^{2p}~d\mu(z)\,$$
is always satisfied.

\subsection{Reduced density matrices}
\label{sec.rdm}
\begin{thm} \cite{AmNi3}. \label{eq.wigcompact}
If the family $(\varrho_{\varepsilon})_{\varepsilon\in \mathcal{E}}$
satisfies $\mathcal{M}(\varrho_{\varepsilon},\varepsilon\in
\mathcal{E})=\left\{\mu\right\}$ with the $(PI)$-condition:
$$
\forall p\in \mathbb{N}\,, \lim_{\varepsilon\in
  \mathcal{E},\varepsilon\to
  0}\Tr\left[\varrho_{\varepsilon}\mathbf{N}^{p}\right]=\int_{\mathcal{Z}}|z|^{2p}~d\mu(z)\;,
$$
then $\Tr\left[\varrho_{\varepsilon}b^{Wick}\right]$ converges to
$\int_{\mathcal{Z}}b(z)~d\mu(z)$ for all polynomial $b(z)$ and
$$
\lim_{\varepsilon\in \mathcal{E}\,, \varepsilon\to 0}\|\gamma_{\varepsilon}^{p}-\gamma_{0}^{p}\|_{\mathcal{L}^{1}}=0
$$
for all $p\in \mathbb N$.
\end{thm}

\begin{thm} \cite{AmNi3,AmNi4,QB}.
Assume  $\mathcal{M}(\varrho_{\varepsilon}, \varepsilon \in
(0,\bar{\varepsilon})=\left\{\mu_{0}\right\}$ 
and the condition $(PI)$\,.
Then
$\mathcal{M}(e^{-i\frac{t}{\varepsilon}H_{\varepsilon}}\varrho_{\varepsilon}e^{i\frac{t}{\varepsilon}H_{\varepsilon}},
\varepsilon
  \in (0,\bar{\varepsilon}))=\left\{\mu_{t}\right\}$. The measure $\mu_{t}=\Phi(t,0)_{*}\mu_{0}$
 is the push-forward measure of the initial measure $\mu_{0}$ where
 $\Phi(t,0)$ is the hamiltonian flow associated with the equation
\begin{equation}\label{eq.Hartree}
  i\partial_{t}z_{k}(t)=-\Delta_{K}z_{k}(t)+\sum_{j}V_{kj}|z_{j}|^{2}z_{k}\,.
\end{equation}
\end{thm} 

After propagation of the Wigner measures, for any $p \in \mathbb N$, the convergence of the reduced densiy matrices is obtained at any time $t$:
$$
\Big\|\gamma_{\varepsilon}^p(t)-\frac{\int_{\mathcal Z}|z^{\otimes
    p}\rangle \langle z^{\otimes p}|d\mu_t(z)}{\int_{\mathcal
     Z}|z|^{2p}d\mu_0(z)}\Big\|_{\mathcal L^1} \longrightarrow 0.
$$

\begin{thm} \cite{AFP}.
\label{th-rate}
Let $(\alpha(n))_{n\in\nz^*}$ be a sequence of positive numbers with $\lim\alpha(n)=\infty$ and such that $(\frac{\alpha(n)}{n})_{n\in\nz^*}$ is bounded.
Let $(\varrho_{n})_{n\in\nz^*}$ and
$(\gamma_{\infty}^{(p)})_{p\in\nz^*}$ be two sequences of density
matrices with $\varrho_n\in\L^1(\bigvee^n \mathcal Z)$ and
$\gamma_\infty^{(p)}\in\L^1(\bigvee^p \mathcal Z)$  for each $n,p\in\nz^*$.  Assume that there exist $C_0>0$, $C>2$ and $\gamma\geq 1$ such that for all
$n,p\in\nz^*$ with $n\geq \gamma p$:
\bea
\label{init-ineq}
\left\|\gamma^{(p)}_{n}-\gamma_{\infty}^{(p)}\right\|_1\leq
 C_0 \frac{C^p}{\alpha(n)}\,.
\eea
 Then for any $T>0$ there exists $C_T>0$ such that for all $t\in[-T,T]$ and all $n,p\in\mathbb{N}^*$ with $n\geq \gamma p$,
\bea
\label{main-ineq}
\left\|\gamma^{(p)}_{n}(t)-\gamma_{\infty}^{(p)}(t)\right\|_1\leq
 C_T \frac{C^p}{\alpha(n)}\,,
\eea
where
\bean
\gamma_{\infty}^{(p)}(t)=\int_{\mathcal Z} |z^{\otimes p}\rangle\langle z^{\otimes p}| \, d\mu_t(z)\,,
\eean
with $\mu_t=(\Phi_t)_\sharp\mu_0$ is the push-forward of the initial measure $\mu_0$ by  the well defined and continuous Hartree  flow $\Phi_t$ on $\mathcal Z$.
\end{thm}

\section{Numerical methods}

\subsection{Method to solve the Hartree equation}
To solve the mean field equation \eqref{eq.Hartree}, a Runge-Kutta method is used.\\

Let $b_i$, $a_{ij}$ $(i,j=1,\ldots,s)$ be real numbers and
$c_i=\sum_{j=1}^sa_{ij}$.
\\
An s-stage Runge-Kutta method with a time step $h$ to solve a first-order ordinary equation
$y'=f(t,y)$ , $y(t_0)=y_0$ is given by:
\begin{eqnarray*}
k_i&=&f(t_0+c_ih,y_0+h\sum_{j=1}^sa_{ij}k_j) \; , i=1,\ldots,s\\
y_1&=&y_0+h\sum_{i=1}^sb_ik_i
\end{eqnarray*}
represented as:

\begin{tabular}{c|ccc}
$c_1$    & $a_{11}$ & \ldots & $a_{1s}$ \\
\vdots & \vdots &        & \vdots \\
$c_s$    & $a_{s1}$ & $\ldots$ & $a_{ss}$ \\
\hline
       & $b_1$   & \ldots &  $b_s$   \\
\end{tabular} .
\\

 Here the system is autonomous, and according to \cite{HLW} the coefficients used for the Gauss RK
 method are:\\
\begin{tabular}{c|cccc}
0   &     &     &     &     \\
1/2 & 1/2 &     &     &     \\
1/2 & 0   & 1/2 &     &     \\
1   & 0   & 0   & 1   &     \\
\hline
    & 1/6 & 2/6 & 2/6 & 1/6 \\
\end{tabular},
\begin{tabular}{c|cccc}
0   &     &     &     &     \\
1/3 & 1/3 &     &     &     \\
2/3 &-1/3 & 1   &     &     \\
1   & 1   & -1  & 1   &     \\
\hline
    & 1/8 & 3/8 & 3/8 & 1/8 \\
\end{tabular}
or
\begin{tabular}{c|cc}
$1/2-\sqrt{3}/6$ & 1/4            & $1/4-\sqrt{3}/6$ \\
$1/2+\sqrt{3}/6$ & $1/4+\sqrt{3}/6$ & 1/4            \\
\hline
               & 1/2            & 1/2            \\
\end{tabular} .

In our case, the function $f$ corresponds to $f(z)=-\mathrm i
\left (-\Delta_{K}z_{k}+\sum_{j}V_{kj}|z_{j}|^{2}z_{k}\right )$.\\
As a function in $\mathbb R^{2K}$ by replacing $z$ by $q+\mathrm i p$,
$$f(q,p)=f_q(q,p)+\mathrm i f_p(q,p)$$
$$f_{q_i}(q,p)= -(p_{i+1}+p_{i-1})+\sum_{j\ne i} V_{ij} (q_j^2+p_j^2)p_i$$
$$f_{p_i}(q,p)=q_{i+1}+q_{i-1}-\sum_{j\ne i} V_{ij}
(q_j^2+p_j^2)q_i$$

For an implicit Runge-Kutta method, a Newton method is applied to find the coefficients $k_i$ for each step of the RK method to the function
$g_{y_0}: (k_i)_{i=1,\cdots s} \mapsto
\left(k_i-f(y_0+h\sum_{j=1}^sa_{ij}k_j)\right)_{i=1,\cdots s}$.\\

Given the time step $h$ small enough, the starting point of the Newton method
is chosen by setting $k_i=f(y_0)$ for all $i$.\\

To apply the Newton's method the differential of $f$ is computed by
using the following partial derivatives of $f$:
\begin{align*}\label{diff.f}
\frac{\partial f_{q_i}}{\partial q_k}&=(1-\delta_{i,k})2V_{ik}q_kp_i\\
\frac{\partial f_{p_i}}{\partial q_k}&=\delta_{i+1,k}+\delta_{i-1,k}+(\delta_{i,k}-1)2V_{ik}q_kq_i-\delta_{i,k}\sum_{j\ne i} V_{ij}(q_j^2+p_j^2)\\
\frac{\partial f_{q_i}}{\partial p_k}&=-(\delta_{i+1,k}+\delta_{i-1,k})+(1-\delta_{i,k})2V_{ik}p_kp_i+\delta_{i,k}\sum_{j\ne i} V_{ij}(q_j^2+p_j^2)\\
\frac{\partial f_{p_i}}{\partial p_k}&=(1-\delta_{i,k})2V_{ik}p_kq_i
\end{align*}
\\

Then the differential of $g_{y_0}$ is:
\label{diff.g}
\begin{align*}
  Dg_{y_0}((k_i)_{i=1,\cdots,s})=\left(Id_{2K}-ha_{il}Df(y_0+h\sum_{j=1}^sa_{ij}k_j)\right)_{i,l=1,\cdots,s}
\end{align*}
where $g_{y_0}$ is considered as a function from $\mathbb R^{2Ks}$.

\subsection{Resolution of the Schr\"odinger equation in $\bigvee^N\mathcal Z$}

\subsubsection{Composition method}
For a given $\Psi$ in $\bigvee^{N}\mathcal Z$, the full $N$-body
evolved state $e^{-i\frac{t}{\varepsilon}H_{\varepsilon}}\Psi$ is
computed in the basis $(e_{\alpha})_{|\alpha|=N}$. After writing
$\Psi=\sum_{|\alpha|=N}\Psi_{\alpha}e_{\alpha}$ a modified splitting
method for which the numerical error is carefully controlled (see
\ref{sec.error-estim}), involves only multiplications by the diagonal matrix
$e^{-i\frac{t}{\varepsilon p}\mathcal V}$ and the sparse matrix $d\Gamma(-\Delta_K)$.\\
In order to handle the high complexity of the problem (see table
\ref{dimensions}) no matrix, but only vectors or the sparse
matrices $d\Gamma(-\Delta_K)$ and the matrix $(V_{ij})_{i,j}$ are stored.\\ 
The complete evolution $e^{-i\frac{t}{\varepsilon}H_{\varepsilon}}$ is computed by a
composition method based on the Strang
splitting method:

$$e^{-i\frac{t}{\varepsilon}H_{\varepsilon}}=\lim_{p\rightarrow
  \infty}\big(e^{-i\frac{t}{2\varepsilon p}\mathcal V}
e^{-i\frac{t}{\varepsilon p}H_0}e^{-i\frac{t}{2\varepsilon p}\mathcal V}\big )^{p} \;\;\;
\;.
$$
The $4^{th}$ order composition method is given by:
\begin{align*}
e^{-i\frac{t}{\varepsilon}H_{\varepsilon}}
=\lim_{p\rightarrow
  \infty}\big(e^{-i\frac{a_{3} t}{2\varepsilon p}\mathcal V}
e^{-i\frac{a_{3} t}{\varepsilon p}H_0}e^{-i\frac{a_{3} t}{2\varepsilon p}\mathcal V}e^{-i\frac{a_{2} t}{2\varepsilon p}\mathcal V}
e^{-i\frac{a_{2} t}{\varepsilon p}H_0}e^{-i\frac{a_{2} t}{2\varepsilon
    p}\mathcal V}
e^{-i\frac{a_{1} t}{2\varepsilon p}\mathcal V}e^{-i\frac{a_{1} t}{\varepsilon p}H_0}e^{-i\frac{a_{1} t}{2\varepsilon
    p}\mathcal V}\big )^{p} \;,
\end{align*}

where the coefficients of the method are satisfying the two
equations (see \cite{HLW}):

\label{eq.composition}
\begin{align}
  a_{1}+a_{2}+a_{3}&=1\\
  a_{1}^{3}+a_{2}^{3}+a_{3}^{3}&=0
\end{align}

and are given by:
\begin{align} \label{coeff.composition}
a_{1}=a_{3}=\frac{1}{2-2^{1/3}} \;,\qquad
a_{2}=-\frac{2^{1/3}}{2-2^{1/3}} \;.
\end{align}\\

 
\subsubsection{Computation of the free evolution $e^{-i\frac{t}{\varepsilon}
  d\Gamma(-\Delta_{K})}$}
The numerical computation of  $e^{-i\frac{t}{\varepsilon}
  d\Gamma(-\Delta_{K})}=\Gamma(e^{it\Delta_{K}})$, relies on the
following two remarks:
\begin{itemize}
\item  the dimension of the $N$-fold sectors $\binom{N+K-1}{K-1}$ prevents
the storage of any square matrix.
\item  the matrix of $\Gamma(e^{it\Delta_{K}})$ is actually non trivial
sparse matrix in the basis $(e_{\alpha})$.
\end{itemize}

 The matrix of $\Delta_K$ is given by:
$$
\Delta_K=
\left(\begin{array}{cccccc}
 0&1& 0 &\cdots&0&1\\
1&\ddots&\ddots&\ddots&&0\\
0&\ddots&\ddots&\ddots& \ddots&\vdots\\
\vdots&\ddots&\ddots&\ddots&\ddots&0\\
0&&\ddots&\ddots&\ddots&1\\
1&0&\cdots&0&1&0 \end{array} \right)\;,
$$\\


We are interested in the matrix of the second quantization of the
discrete Laplacian on the basis of the bosons space to implement it
numerically as a sparse matrix containing only $2K{N+K-2 \choose K-2}$ elements
whereas a full matrix contains  ${N+K-1 \choose K-1}^2$.\\
Then $e^{-i\frac{\Delta t}{\varepsilon}
  d\Gamma(-\Delta_{K})}$ will be computed at each time step by a
$4^{th}$ order Taylor expansion.\\
This expansion is then replaced in the composition method.\\

For an operator $A:\mathbb{C}^{K}\longrightarrow \mathbb{C}^{K}$, $A=(A_{i,j})_{i,j}\;,$
$$
d\Gamma(A)_{|\bigvee^n  \mathbb{C}^{K}}=\sum_{i,j=1}^K
A_{i,j} a^*(e_i) a(e_j)\;.
$$
This yields:
\begin{eqnarray}
d\Gamma(-\Delta_K)&=&-\sum_{j=1}^K  a^*(e_{j+1}) a(e_j)+a^*(e_j)
a(e_{j+1})\;. \label{laplacien}
\end{eqnarray}

\begin{lem} \label{lem.deriv-annih}
For all multi-indices $\gamma$ and $\alpha$ the following equality holds:\\
$$
a(e)^{\gamma}a^*(e)^{\alpha}|\Omega\rangle=\delta_{\gamma\leq\alpha}
\varepsilon^{|\gamma|}\frac{\alpha!}{(\alpha-\gamma)!}a^*(e)^{\alpha-\gamma}|\Omega\rangle\;.
$$
\end{lem}

 \proof

\begin{eqnarray*}
a(e)^{\gamma}a^*(e)^{\alpha}&=&a(e_1)^{\gamma_1}\ldots a(e_K)^{\gamma_K}a^*(e_1)^{\alpha_1}\ldots a^*(e_K)^{\alpha_K}\\
&=&\prod_{i=1}^{K}a(e_i)^{\gamma_i}a^*(e_i)^{\alpha_i}
\mbox{ which is a commutative product because of CCR \eqref{CCR}}\\
&=&
\bigotimes_{i=1}^{K} a(e_i)^{\gamma_i}a^*(e_i)^{\alpha_i} 
\end{eqnarray*}

by using the following separation of the variables:
$
\Gamma(\mathcal Z)=\Gamma(\mathbb C e_{1})\otimes\ldots \otimes
\Gamma(\mathbb C e_{K}) \;.
$\\

In this space let $|\Omega\rangle$ be $|\Omega_{1}\rangle\otimes
\ldots \otimes |\Omega_{K}\rangle$.\\

Let us consider $\gamma_{i}\geq 1$ and $\alpha_{i}\geq 1$,
\begin{align*}
a(e_i)^{\gamma_i}a^*(e_i)^{\alpha_i}|\Omega_{i}\rangle&=a(e_i)^{\gamma_i-1}a(e_i)a^*(e_i)^{\alpha_i}|\Omega_{i}\rangle\\
&=a(e_i)^{\gamma_i-1}a^*(e_i)^{\alpha_i}a(e_i)|\Omega_{i}\rangle+a(e_i)^{\gamma_i-1}[a(e_i),\,a^*(e_i)^{\alpha_i}]|\Omega_{i}\rangle\\
&=\varepsilon
\alpha_{i}a(e_i)^{\gamma_i-1}a^*(e_i)^{\alpha_i-1}|\Omega_{i}\rangle \;.
\end{align*}\\

By induction, we obtain
$$
a(e_i)^{\gamma_i}a^*(e_i)^{\alpha_i}|\Omega_{i}\rangle=\alpha_{i}\varepsilon
\times(\alpha_{i}-1)\varepsilon\times \ldots \times 2 \varepsilon
\times \varepsilon a(e_i)^{\gamma_i-\alpha_{i}}|\Omega_{i}\rangle=0
\;,  \mbox{when } \alpha_{i}< \gamma_{i}
$$\\

and
\begin{align*}
a(e_i)^{\gamma_i}a^*(e_i)^{\alpha_i}|\Omega_{i}\rangle&=\alpha_{i}\varepsilon
\times(\alpha_{i}-1)\varepsilon\times \ldots \times
(\alpha_{i}-(\gamma_{i}-1))\varepsilon
a^{*}(e_i)^{\alpha_i-\gamma_{i}}|\Omega_{i}\rangle\\
&=\varepsilon^{\gamma_{i}}\frac{\alpha_{i}!}{(\alpha_{i}-\gamma_{i})!}a^{*}(e_i)^{\alpha_i-\gamma_{i}}|\Omega_{i}\rangle
\;, \mbox{when  }\gamma_{i} \leq \alpha_{i}.
\end{align*}\\

The above separation of variables leads under the condition $\gamma
\leq\alpha$ to:
\begin{align*}
a(e)^{\gamma}a^*(e)^{\alpha}|\Omega\rangle&=\otimes_{i=1}^{K}
a(e_i)^{\gamma_i}a^*(e_i)^{\alpha_i}|\Omega_{i}\rangle\\
&=\Big(\prod_{i=1}^{K}\frac{\varepsilon^{\gamma_{i}}\alpha_{i}!}{(\alpha_{i}-\gamma_{i})!}\Big) \otimes_{i=1}^{K}
a^*(e_i)^{\alpha_i-\gamma_{i}}|\Omega_{i}\rangle\\
&=\varepsilon^{|\gamma|}\frac{\alpha!}{(\alpha-\gamma)!}\Big(\prod_{i=1}^{K}a^*(e_i)^{\alpha_i-\gamma_{i}}\Big)(|\Omega_{1}\rangle\otimes
\ldots \otimes |\Omega_{K}\rangle)\\
&=\varepsilon^{|\gamma|}\frac{\alpha!}{(\alpha-\gamma)!}a^*(e)^{\alpha-\gamma}|\Omega\rangle \;.
\end{align*}

\fin

\begin{prop} 
For all multi-indices $\alpha$ and $\beta$, the matrix elements of $d\Gamma(-\Delta_{K})$ are given by:
$$
d\Gamma(-\Delta_{K})_{\alpha,\beta}=-\varepsilon \sum_{i}(\delta^{+}_{\beta-e_{i},\alpha-e_{i+1}}\sqrt{\beta_{i}(\beta_{i+1}+1)}+\delta^{+}_{\beta-e_{i+1},\alpha-e_{i}}\sqrt{\beta_{i+1}(\beta_{i}+1)})\;,
$$

where $\delta^{+}_{\alpha,\beta}=\delta_{\alpha,\beta}{\bf 1}_{\mathbb N^{K}}(\alpha)$ for $\alpha$ and $\beta$
multi-indices in $\mathbb Z^{K}$.
\end{prop}

\proof

According to \eqref{laplacien} and to Lemma
\ref{lem.deriv-annih}, we obtain:\\

\begin{align*}
a^*(e_{i+1})a(e_i) a^*(e)^{\alpha}|\Omega\rangle=&\delta_{e_{i}\leq\alpha}
\varepsilon\frac{\alpha!}{(\alpha-e_{i})!}a^*(e)^{e_{i+1}+\alpha-e_{i}}|\Omega\rangle\\
=&\delta_{1\leq\alpha_{i}}
\varepsilon\alpha_{i}a^*(e)^{e_{i+1}+\alpha-e_{i}}|\Omega\rangle \;,\\
\end{align*}

and
\begin{align*}
a^*(e_{i})a(e_{i+1}) a^*(e)^{\alpha}|\Omega\rangle=&\delta_{e_{i+1}\leq\alpha}
\varepsilon\frac{\alpha!}{(\alpha-e_{i+1})!}a^*(e)^{e_{i}+\alpha-e_{i+1}}|\Omega\rangle\\
=&\delta_{1\leq\alpha_{i+1}}
\varepsilon\alpha_{i+1}a^*(e)^{e_{i}+\alpha-e_{i+1}}|\Omega\rangle \;.\\
\end{align*}\\

Then
\begin{align*}
\langle e_{\alpha}, d\Gamma(-\Delta_{K}) e_{\beta}
\rangle &=-\left\langle e_{\alpha},\left (\sum_{i=1}^K  a^*(e_{i+1})
  a(e_i)+a^*(e_i) a(e_{i+1}) \right ) e_{\beta}
\right \rangle\\
&=\frac{-\varepsilon}{\sqrt{\alpha!\beta!\varepsilon^{2N}}}\sum_{i}\left\langle
a^{*}(e)^{\alpha}\Omega|\left (\delta_{1\leq\beta_{i}}\beta_{i}
a^{*}(e)^{\beta+e_{i+1}-e_{i}}+\delta_{1\leq\beta_{i+1}}\beta_{i+1}
a^{*}(e)^{\beta+e_{i}-e_{i+1}} \right)\Omega \right\rangle\\
&=\frac{-\varepsilon}{\varepsilon^{N}\sqrt{\alpha!\beta!}}\sum_{i}
(\delta^{+}_{\beta-e_{i},\alpha-e_{i+1}}\beta_{i}\varepsilon^{N}\sqrt{\alpha!(\beta-e_{i}+e_{i+1})!}\\
&+\delta^{+}_{\beta-e_{i+1},\alpha-e_{i}}\beta_{i+1}\varepsilon^{N}\sqrt{\alpha!(\beta -e_{i+1}+e_{i})!})\\
&=\frac{-\varepsilon}{\sqrt{\beta!}}\sum_{i}
(\delta^{+}_{\beta-e_{i},\alpha-e_{i+1}}\beta_{i}\sqrt{(\beta -e_{i}+e_{i+1})!}+\delta^{+}_{\beta-e_{i+1},\alpha-e_{i}}\beta_{i+1}\sqrt{(\beta
  -e_{i+1}+e_{i})!})\\
&=-\varepsilon \sum_{i}
(\delta^{+}_{\beta-e_{i},\alpha-e_{i+1}}\beta_{i}\sqrt{\frac{\beta_{i+1}+1}{\beta_{i}}}+\delta^{+}_{\beta-e_{i+1},\alpha-e_{i}}\beta_{i+1}\sqrt{\frac{\beta_{i}+1}{\beta_{i+1}}})\\
d\Gamma(-\Delta_{K})_{\alpha,\beta}&=-\varepsilon \sum_{i}
(\delta^{+}_{\beta-e_{i},\alpha-e_{i+1}}\sqrt{\beta_{i}(\beta_{i+1}+1)}+\delta^{+}_{\beta-e_{i+1},\alpha-e_{i}}\sqrt{\beta_{i+1}(\beta_{i}+1)})
\;.
\end{align*}\\

\fin 

Numerically only the indices of the
multi-indices $\alpha$ and $\beta$ corresponding to the nonzero components
of $d\Gamma(-\Delta_K)$ with their values, are stored in an array.\\
In the algorithm instead of running over the multi-indices $\alpha$ or
$\beta$ with a length $N$, the multi-indices $\beta'$ with
a length $N-1$ are run over. And for each $i$ in $[1,K]$, the changes
of multi-indices
$\beta'=\beta-e_i$ or $\beta'=\beta-e_{i+1}$ are used, then the indices
of the corresponding multi-indices $\alpha$ and $\beta$ with length
$N$ are looked for.\\
Therefore an array composed of
$2K\binom{N+K-2}{K-2}$ triplets of elements is numerically stored.\\


\subsubsection{Computation of the interaction factor $e^{-i\frac{t}{\varepsilon}\mathcal V}$}

Denote $a^{\#}_j=a^{\#}(e_j)$ and $N_i=a^*_ia_i$ .\\

By using the relations CCR \eqref{CCR}:
\begin{align*}
a^*_ia^*_ja_ia_j&=a^*_i(a_ia^*_j-\varepsilon\delta_{ij})a_j=a^*_ia_ia^*_ja_j-\varepsilon\delta_{ij}a^*_ia_j\\
&=N_iN_j-\varepsilon\delta_{ij}N_i=N_i(N_j-\varepsilon\delta_{ij}) \;.
\end{align*}

Then $\mathcal V$ can be rewritten as:
$$
\mathcal{V}=\frac{1}{2}\sum_{(i,j)\in\mathbb{Z}_{K}^{2}}V_{ij}N_i(N_j-\varepsilon\delta_{ij})\;.
$$\\

And since $N_ie_\alpha=\varepsilon\alpha_ie_\alpha$ then
$\mathcal V$ is diagonal in the basis $(e_{\alpha})_{\alpha}$:
\begin{align*}
\mathcal{V}e_\alpha&=\left(\frac{1}{2}\sum_{(i,j)\in\mathbb{Z}_{K}^{2}}V_{ij}\varepsilon\alpha_i(\varepsilon\alpha_j-\varepsilon\delta_{ij})\right)e_\alpha\\
&=\left(\frac{\varepsilon^{2}}{2}\sum_{(i,j)\in\mathbb{Z}_{K}^{2}}V_{ij}\alpha_i(\alpha_j-\delta_{ij})\right)e_\alpha
\end{align*}

and
\begin{eqnarray*} e^{-i\frac{t}{\varepsilon}\mathcal V}e_\alpha &=&
  e^{-i\frac{t}{\varepsilon}\left(\frac{\varepsilon^2}{2}\sum_{(i,j)\in\mathbb{Z}_{K}^{2}}V_{ij}\alpha_i(\alpha_j-\delta_{ij})\right)}e_\alpha \\
&=& e^{-it\frac{\varepsilon}{2}\left(\sum_{i\neq j}V_{ij}\alpha_i\alpha_j+\sum_{i\in\mathbb{Z}_{K}}V_{ii}\alpha_i(\alpha_i-1)\right)}e_\alpha\;.
\end{eqnarray*}

\subsection{Numerical computation of the reduced density matrices}
Consider $b^{Wick}=a^*(e)^{\delta}a(e)^{\gamma}$ with
$|\delta|=|\gamma|$ and its associated homogeneous polynomial:
$$
b(z)=\bar z^{\delta}z^{\gamma}=\bar z_1^{\delta_1}\ldots \bar
z_K^{\delta_K} z_1^{\gamma_1}\ldots z_K^{\gamma_K} \;.
$$

 Let us compute the quantity
$\mathrm{Tr}(\varrho_{\varepsilon}b^{Wick})$ when $\varrho_{\varepsilon}$ is a
normal state. Using an orthonormal basis of the $N$-fold sector
$\bigvee^N \mathcal Z$, $\varrho_{\varepsilon}$ is a linear combination
of operators $|\Phi
\rangle \langle \Psi|$. It suffices to compute $\mathrm{Tr}(|\Phi
\rangle \langle \Psi|b^{Wick})=\langle
\Psi , b^{Wick}\Phi\rangle$.
\\
\begin{lem}\label{lem.trace}
Set $b^{Wick}=a^*(e)^{\delta}a(e)^{\gamma}$ with
$|\delta|=|\gamma|$ and let $\Phi$ and $\Psi$ be in $\bigvee^N \mathcal Z$ then:

$$
\langle\Psi , b^{Wick}\Phi\rangle=\varepsilon^{|\gamma|}\sum_{|\alpha'|=N-|\delta|}\bar
\Psi_{\alpha'+\delta}\Phi_{\alpha'+\gamma}\frac{\sqrt{(\alpha'+\delta)!(\alpha'+\gamma)!}}{\alpha'!} 
$$
in the orthonormal basis
$\left(\frac{a^*(e)^{\alpha}}{\sqrt{\varepsilon^{N}\alpha!}}|\Omega\rangle\right)_{|\alpha|=N}$ of $\bigvee^N \mathcal Z$.
\end{lem}

\proof
 Given $\Psi$ and $\Phi$ in the N-particles bosons
 space and the formula \ref{lem.deriv-annih}
$$
a^*(e)^{\delta}a(e)^{\gamma}a^*(e)^{\alpha}|\Omega\rangle=\delta_{\gamma\leq\alpha}\varepsilon^{|\gamma|}\frac{\alpha!}{(\alpha-\gamma)!}a^*(e)^{\delta+\alpha-\gamma}|\Omega\rangle\;,
$$

 we can write :
\begin{align*}
\langle\Psi , b^{Wick}\Phi\rangle&=\langle
\sum_{|\alpha|=N}\Psi_{\alpha}\frac{a^*(e)^{\alpha}}{\sqrt{\varepsilon^{N}\alpha!}}\Omega\;,\;
 \varepsilon^{|\gamma|}\sum_{|\beta|=N,\beta\geq\gamma}\Phi_{\beta}\frac{\sqrt{\beta!}}{\varepsilon^{N/2}(\beta-\gamma)!}a^*(e)^{\delta+\beta-\gamma}|\Omega\rangle&\\
&=\frac{\varepsilon^{|\gamma|}}{\varepsilon^{N}}\sum_{|\alpha|=N}\bar
\Psi_{\alpha}\sum_{|\beta|=N,\beta\geq\gamma}\Phi_{\beta}\frac{\sqrt{\beta!}}{\sqrt{\alpha!}(\beta-\gamma)!}
\langle a^*(e)^{\alpha}\Omega,a^*(e)^{\delta+\beta-\gamma}|\Omega\rangle&\\
&=\varepsilon^{|\gamma|}\sum_{|\alpha|=N,\alpha\geq\delta}\bar \Psi_{\alpha}\Phi_{\alpha+\gamma-\delta}\frac{\sqrt{(\alpha+\gamma-\delta)!}}{\sqrt{\alpha!}(\alpha-\delta)!}\alpha!&\\
&=\varepsilon^{|\gamma|}\sum_{|\alpha|=N,\alpha\geq\delta}\bar \Psi_{\alpha}\Phi_{\alpha+\gamma-\delta}\frac{\sqrt{\alpha!(\alpha+\gamma-\delta)!}}{(\alpha-\delta)!}&\\
&=\varepsilon^{|\gamma|}\sum_{|\alpha'|=N-|\delta|}\bar
\Psi_{\alpha'+\delta}\Phi_{\alpha'+\gamma}\frac{\sqrt{(\alpha'+\delta)!(\alpha'+\gamma)!}}{\alpha'!}  \;.
\end{align*}
\\
because
$$
\langle a^*(e)^{\alpha}\Omega,a^*(e)^{\delta+\beta-\gamma}|\Omega\rangle
\neq 0
$$
if and only if $\alpha=\delta+\beta-\gamma$ so
$\beta=\alpha+\gamma-\delta$ and $\beta \geq \gamma$ means
$\alpha-\delta \geq 0$. 
\\
The last line is obtained by a change of multi-indices by setting for each $\alpha$, $\alpha'=\alpha-\delta$
because $\alpha \geq \delta$, and then $|\alpha'|=|\alpha|-|\delta|=N-|\delta|$.
\fin
\\

Numerically, all multi-indices of $\mathbb
N^K$ with length not larger than a given $N_{max}$ are stored in the lexicographic order.
\\
For our algorithms, we pay attention to preserve this lexicographic
order (or reverse).
\\
For a given $N \leq N_{max}$, the list of relevant multi-indices (with
length $N$) is
extracted and handled in the lexicographic
order.
\\

For a given $\delta$, numerically the above summation is performed over
multi-indices $\alpha'$ such that $|\alpha'|=N-|\delta|$ in the
lexicographic order. 
\\
Then for each $\alpha'$, the multi-indices $\alpha$ of length
$N$ written as $\alpha=\alpha'+\delta$ are looked for. These $\alpha$
are exactly the multi-indices such that $\alpha \geq \delta$ and
$|\alpha|=N$.\\
Note in particular that the mapping $\alpha' \mapsto
\alpha'+\delta$ preserves the lexicographic order.
\\

First let us compute the matrix elements of $\gamma_{\varepsilon}^p$ in the orthonormal
basis $(e_{\alpha})_{\alpha} $.\\

 The matrix element corresponding to the line $\beta$ and
column $\alpha$ is:

\begin{align*}
\Big\langle
\frac{a^*(e)^{\beta}}{\sqrt{\varepsilon^p\beta!}}\Omega\Big|\gamma_{\varepsilon}^p
\frac{a^*(e)^{\alpha}}{\sqrt{\varepsilon^p\alpha!}}\Big|\Omega\Big\rangle
&=\Tr\Big(\gamma_{\varepsilon}^p(t)\Big|\frac{a^*(e)^{\alpha}}{\sqrt{\varepsilon^p\alpha!}}\Omega\Big\rangle\Big\langle
\frac{a^*(e)^{\beta}}{\sqrt{\varepsilon^p\beta!}}\Omega\Big|\Big)\\
&=\frac{\Tr(\varrho_\varepsilon(t) b^{Wick})}{\varepsilon^p N(N-1)\ldots(N-p+1)}
\end{align*}

according to the duality relation \eqref{rdm-duality} of the reduced
density matrices with $\tilde b=\Big|\frac{a^*(e)^{\alpha}}{\sqrt{\varepsilon^p\alpha!}}\Omega\Big\rangle\Big\langle
\frac{a^*(e)^{\beta}}{\sqrt{\varepsilon^p\beta!}}\Omega\Big|$, and $
b(z)=\langle z^{\otimes p},\tilde b z^{\otimes p}\rangle \in \mathcal
P_{p,p} \;.
$\\

If $\tilde b=\Big|\frac{a^*(e)^{\alpha}}{\sqrt{\varepsilon^p\alpha!}}\Omega\Big\rangle\Big\langle
\frac{a^*(e)^{\beta}}{\sqrt{\varepsilon^p\beta!}}\Omega\Big|$, its Wick quantized is :
$$
b^{Wick}=\frac{p!}{\sqrt{\alpha!\beta!}}a^*(e)^{\alpha}a(e)^{\beta} \;.
$$

And then
$$
\gamma_{\varepsilon}^p(t)(\beta,\alpha)=\frac{p!}{\sqrt{\alpha!\beta!}}\frac{\Tr(\varrho_{\varepsilon}(t)a^*(e)^{\alpha}a(e)^{\beta})}{\varepsilon^p
  N(N-1)\ldots(N-p+1)} \;.
$$

Then owing to Lemma \ref{lem.trace}, all the elements of the matrices
$\gamma_{\varepsilon}^p$ can be numerically computed.
\\

In the case where the initial state is a Hermite state
$\varrho_{\varepsilon}=|z^{\otimes N}\rangle\langle z^{\otimes N}|$, $z^{\otimes N}$ needs to be expanded in the orthonormal basis $(e_{\alpha})$
which is given by the following lemma.

\begin{lem}\label{Hermite-state}
For all $p \in \mathbb N$, and $z \in \mathcal Z$, we obtain in the basis  $(e_{\alpha})_{\alpha}$:
$$
z^{\otimes p}=\sum_{|\alpha|=p}\sqrt{\frac{p!}{\alpha!}}z^{\alpha}e_{\alpha}\;.
$$
\end{lem}

 \proof
\begin{align*}
a^*(z)^p|\Omega\rangle=a^*(z_1e_1+\ldots+z_Ke_K)^p|\Omega\rangle
&=(z_1a^*(e_1)+\ldots+z_Ka^*(e_K))^p|\Omega\rangle&\\
&=\sum_{|\alpha|=p}\frac{p!}{\alpha!}z_1^{\alpha_1}\ldots z_K^{\alpha_K}a^*(e_1)^{\alpha_1}\ldots a^*(e_K)^{\alpha_K}|\Omega\rangle&\\
&=\sum_{|\alpha|=p}\frac{p!}{\alpha!}z^{\alpha}a^*(e)^{\alpha}|\Omega\rangle \;.
\end{align*}

 And then
$$
z^{\otimes p}=\frac{a^*(z)^p}{\sqrt{\varepsilon^p
    p!}}|\Omega\rangle=\sum_{|\alpha|=p}\sqrt{\frac{p!}{\alpha!}}z^{\alpha}e_{\alpha}\;.
$$
\fin

In the case where the initial state is a twin state, the following
lemma is used to obtain an expansion in the basis $(e_{\alpha})$.

\begin{lem}\label{Twin-state}
Let  $\phi \; ,\psi$ be in $\mathcal Z$ and $z$ the state

 $$
\frac{a^{*}(\phi)^{n}a^{*}(\psi)^{m}}{\sqrt{\varepsilon^{n+m}n!m!}}|\Omega\rangle
$$
such that $n+m=N$.

Then we obtain
$$
z=\sqrt{n!m!}\sum_{|\gamma|=N}\Big(\sum_{|\alpha|=n,\alpha \leq  \gamma}
\frac{\sqrt{\gamma!}}{\alpha!(\gamma-\alpha)!}\phi^{\alpha}\psi^{\gamma-\alpha}\Big)
\frac{a^*(e)^{\gamma}}{\sqrt{\varepsilon^N\gamma!}}|\Omega \rangle \;.
$$
\end{lem}

\proof
$$
a^{*}(\phi)^{n}=\sum_{|\alpha|=n}\frac{n!}{\alpha!}\phi^{\alpha}a^*(e)^{\alpha}
$$
$$
a^{*}(\psi)^{m}=\sum_{|\beta|=m}\frac{m!}{\beta!}\psi^{\beta}a^*(e)^{\beta}
$$
$$
a^{*}(\phi)^{n}a^{*}(\psi)^{m}=\sum_{|\alpha|=n,|\beta|=m}\frac{n!m!}{\alpha!\beta!}\phi^{\alpha}\psi^{\beta}
a^*(e)^{\alpha}a^*(e)^{\beta}=\sum_{|\alpha|=n,|\beta|=m}\frac{n!m!}{\alpha!\beta!}\phi^{\alpha}\psi^{\beta}
a^*(e)^{\alpha+\beta}
$$
because of the CCR relations \eqref{CCR}.

\begin{align*}
\frac{a^{*}(\phi)^{n}a^{*}(\psi)^{m}}{\sqrt{n!m!}}|\Omega\rangle&=\sum_{|\alpha|=n,|\beta|=m}\sqrt{n!m!}\frac{\sqrt{(\alpha+\beta)!}}{\alpha!\beta!}\phi^{\alpha}\psi^{\beta}
\frac{a^*(e)^{\alpha+\beta}}{\sqrt{(\alpha+\beta)!}}|\Omega\rangle\\
&=\sqrt{n!m!}\sum_{|\gamma|=N,|\alpha|=n,\alpha \leq \gamma}
\frac{\sqrt{\gamma!}}{\alpha!(\gamma-\alpha)!}\phi^{\alpha}\psi^{\gamma-\alpha}
\frac{a^*(e)^{\gamma}}{\sqrt{\gamma!}}|\Omega \rangle \;.
\end{align*}
By setting $\gamma=\alpha+\beta$,

$$
z=\sqrt{n!m!}\sum_{|\gamma|=N}\Big(\sum_{|\alpha|=n,\alpha \leq  \gamma}
\frac{\sqrt{\gamma!}}{\alpha!(\gamma-\alpha)!}\phi^{\alpha}\psi^{\gamma-\alpha}\Big)
\frac{a^*(e)^{\gamma}}{\sqrt{\varepsilon^N\gamma!}}|\Omega \rangle \;.
$$
\fin

Further the limit reduced density
matrices \eqref{rdm-limit} have to be computed numerically. In order to do this, the integration over
$\mathcal Z$ of the Wigner measure is discretized. The problem is then
reduced to the computation of the matrix elements of $|z^{\otimes
  p}\rangle \langle z^{\otimes p}|$ in the basis $(e_{\alpha})_{|\alpha|=p}$. 
\\

Compute the matrix elements of $|z^{\otimes p}\rangle\langle
z^{\otimes p}|$, according to Lemma \ref{Hermite-state}:
$$
\langle z^{\otimes p}|\frac{a^*(e)^{\alpha}}{\sqrt{\varepsilon^p\alpha!}}|\Omega\rangle=\sqrt{\frac{p!}{\alpha!}}\bar
z^{\alpha}
$$

Then
\begin{align*}
(|z^{\otimes p}\rangle \langle z^{\otimes
  p}|)e_{\alpha}
&=\sqrt{\frac{p!}{\alpha!}}\bar z^{\alpha}\sum_{|\beta|=p}\sqrt{\frac{p!}{\beta!}}z^{\beta}e_{\beta}\\
&=\sum_{|\beta|=p}\frac{p!}{\sqrt{\alpha!\beta!}}\bar
z^{\alpha}z^{\beta}e_{\beta} \;.
\end{align*}

For the computation of the integral $\int_{\mathcal Z}|z^{\otimes p}\rangle \langle z^{\otimes p}|d\mu(z)$, the Wigner measure is
  approximated by a convex combination of gauge invariant delta functions  $\delta_z^{S^1}$, where $\delta_z^{S^1}=\frac 1{2\pi}\int_0^{2\pi}\delta_{e^{\mathrm i \theta}z}d\theta$.\\

For the Wigner measure associated with the Hermite states,
$\mu=\delta_z^{S^1}$, and the discretization is trivial and exact.
it is not needed to be approximated because of the gauge invariance.
\\
In the case of the twin states given by
$\Psi_N=\frac{a^*(\psi_1)^{n_1}a^*(\psi_2)^{n_2}}{\sqrt{\varepsilon^{n_1+n_2}n_1!n_2!}}|\Omega\rangle$,
where $\psi_1 \;, \psi_2 \in \mathcal Z$,
$\|\psi_1\|=\|\psi_2\|=1$, and $n_1=n_2=\frac{N}2$, the Wigner measure is
$\mu_0=\frac1{2\pi}\int_0^{2\pi}\delta^{S^1}_{\psi_{\phi}}d\phi$
according to \cite{AmNi4}, with:
$$\psi_{\phi}=\cos(\phi)\psi_0+\sin(\phi)\psi_{\frac{\pi}2}\;,$$ 
$$\psi_0=\frac{\sqrt{2}}{2}(\psi_1+\psi_2) ,\quad
\psi_{\frac{\pi}2}=\mathrm i \frac{\sqrt{2}}2(\psi_1-\psi_2) \;.$$

Numerically the interval $[0,2\pi]$ is discretized and $\mu_0$ is
approximated by $\frac 1m \sum_{k=1}^m\delta_{z_k}^{S^1}$.
\\
The Wigner measure $\mu_t$ propagated at the time $t$ of the
twin states is given by :
$$\mu_t=\frac1{2\pi}\int_0^{2\pi}\delta^{S^1}_{\psi_{\phi}(t)}d\phi
\;,$$
where $\psi_{\phi}(t)$ is solution to the Hartree equation at the time
$t$ with initial condition $\psi_{\phi}$.\\

Numerically it is now approximated by
$\frac{1}{m}\sum_{k=1}^m\delta_{z_k(t)}^{S^1}$ where $z_k(t)$ solves the Hartree equation \eqref{eq.Hartree}.\\

Thus the matrix elements of $\int_{\mathcal Z}|z^{\otimes
    p}\rangle \langle z^{\otimes p}|d\mu_t(z)$ are given by the formula:
 $$\frac{1}{m}\sum_{k=1}^m\frac{p!}{\sqrt{\alpha!\beta!}}\bar
z_{k}(t)^{\alpha}z_{k}(t)^{\beta} \;.
$$
And the approximation of the scalar $\int_{\mathcal
  Z}|z|^{2p}d\mu_0(z)$ is given by the formula $\frac{1}{m}\sum_{k=1}^m|z_k|^{2p}$.\\

 The matrix of
$\gamma_{\varepsilon}^p(t)-\frac{\int_{\mathcal Z}|z^{\otimes
    p}\rangle \langle z^{\otimes p}|d\mu_t(z)}{\int_{\mathcal
    Z}|z|^{2p}d\mu_0(z)}$ can then be computed at any time $t$ numerically with a good approximation.
\\

\section{Error estimates}
\label{sec.error-estim}

\subsection{Error estimate of the composition method}
The Baker-Campbell-Hausdorff formula (see \cite{HLW}) allows to find the
order of the composition method which is $4$. Then the Taylor's formula with $4^{th}$
order integral remainder and the Cauchy inequalities are used to estimate the error.\\
The following proposition gives an estimate of the composition method.\\

\begin{prop} \label{comp estimates}
 Let $R>0$, $A$ and $B$ be two anti-adjoint matrices such that
 $(a_{1}-a_{2})\|A\|-\frac{3a_{2}}{2}\|B\| \leq R$.\\
Then
$$
\|e^{A+B}-\Psi_{A,B}\| \leq \frac{2e^{R}}{R^{5}}\left ((a_{1}-a_{2})\|A\|-\frac{3a_{2}}{2}\|B\|\right)^{5}
$$
where
$\Psi_{A,B}=e^{\frac{a_{1}}{2}B}e^{a_{1}A}e^{\frac{a_{1}}{2}B}e^{\frac{a_{2}}{2}B}e^{a_{2}A}e^{\frac{a_{2}}{2}B}e^{\frac{a_{1}}{2}B}e^{a_{1}A}e^{\frac{a_{1}}{2}B}$
is the composition method.
\end{prop}

\proof
\begin{align*}
e^{A+B}-\Psi_{A,B}&=e^{A+B}-e^{\frac{a_{1}}{2}B}e^{a_{1}A}e^{\frac{a_{1}}{2}B}e^{\frac{a_{2}}{2}B}e^{a_{2}A}e^{\frac{a_{2}}{2}B}e^{\frac{a_{1}}{2}B}e^{a_{1}A}e^{\frac{a_{1}}{2}B}\\
&=e^{A+B}-e^{\frac{a_{1}}{2}B}e^{a_{1}A}e^{\frac{a_{1}+a_{2}}{2}B}e^{a_{2}A}e^{\frac{a_{2}+a_{1}}{2}B}e^{a_{1}A}e^{\frac{a_{1}}{2}B}\\
\end{align*}

\begin{align*}
&\|e^{A+B}-e^{\frac{a_{1}}{2}B}e^{a_{1}A}e^{\frac{a_{1}}{2}B}e^{\frac{a_{2}}{2}B}e^{a_{2}A}e^{\frac{a_{2}}{2}B}e^{\frac{a_{1}}{2}B}e^{a_{1}A}e^{\frac{a_{1}}{2}B}\|\\
&=\|e^{-\frac{a_{1}+a_{2}}{2}B}e^{-a_{1}A}e^{-\frac{a_{1}}{2}B}e^{A+B}-e^{a_{2}A}e^{\frac{a_{2}+a_{1}}{2}B}e^{a_{1}A}e^{\frac{a_{1}}{2}B}\|\\
\end{align*}

Then for $z \in \mathbb C$ 
\begin{align*}
&\|e^{z(A+B)}-\Psi_{zA,zB}\| \\
&\leq e^{-\frac{a_{1}+a_{2}}{2}|Im(z)|\|B\|}e^{a_{1}|Im(z)|
  \|A\|}e^{\frac{a_{1}}{2}|Im(z)|\|B\|}e^{|Im(z)| (\|A\|+\|B\|)}\\
&+e^{-a_{2}|Im(z)|\|A\|}e^{-\frac{a_{2}+a_{1}}{2}|Im(z)|\|B\|}e^{a_{1}|Im(z)|\|A\|}e^{\frac{a_{1}}{2}|Im(z)|\|B\|}\\
&=e^{-\frac{a_{1}+a_{2}}{2}|Im(z)|\|B\|}e^{a_{1}|Im(z)|
  \|A\|}e^{\frac{a_{1}}{2}|Im(z)|\|B\|}(e^{|Im(z)|
  (\|A\|+\|B\|)}+e^{-a_{2}|Im(z)|\|A\|})\\
&=e^{|Im(z)|(-\frac{a_{2}}{2}\|B\|+a_{1}
  \|A\|)}(e^{|Im(z)| (\|A\|+\|B\|)}+e^{-a_{2}|Im(z)|\|A\|})\\
&\leq 2e^{|Im(z)|(-\frac{a_{2}}{2}\|B\|+a_{1}
  \|A\|)}e^{-a_{2}|Im(z)| (\|A\|+\|B\|)}\\
&=2e^{|Im(z)|((a_{1}-a_{2})
  \|A\|-\frac{3a_{2}}{2}\|B\|)} \;.\\
\end{align*}

Let us consider the holomorphic function on $\mathbb C$ defined by:
$$
f_{A,B}(z)=e^{zA+zB}-\Psi_{zA,zB} \;.
$$

Since the composition method is of $4^{th}$ order then
for $\lambda \in \mathbb R$, the Taylor's formula with integral remainder yields:

\begin{align*}
f_{A,B}(\lambda)&=\int_{0}^{\lambda}\frac{(\lambda-t)^{4}}{4!}f_{A,B}^{(5)}(
t)dt\\
\|f_{A,B}(\lambda)\|&\leq \frac{\lambda^{5}}{5!} \sup_{t \in
  [0,\lambda]}\|f_{A,B}^{(5)}(t)\| \;.
\end{align*}

By the Cauchy's integral formula, we know for each $t \in [0,1]$:
\begin{align*}
\Big\|\frac{f_{A,B}^{(5)}(t)}{5!}\Big\|&=\Big\|\frac{1}{2i\pi}\int_{|z-t|=1}\frac{f_{A,B}(z)}{(z-t)^{6}}dz\Big\|\\
& \leq
\frac{1}{2\pi}\int_{0}^{2\pi}\|f_{A,B}(t+e^{i\theta})\|d\theta\\
& \leq \sup_{|Im(z)|\leq 1}\|f_{A,B}(z)\|\\
& \leq \sup_{|Im(z)|\leq 1}  2e^{|Im(z)|((a_{1}-a_{2})
  \|A\|-\frac{3a_{2}}{2}\|B\|)}\\
&\leq 2e^{(a_{1}-a_{2})
  \|A\|-\frac{3a_{2}}{2}\|B\|} \;.
\end{align*}

Hence for all $A_{R}$ and $B_{R}$ such that $(a_{1}-a_{2})\|A_{R}\|-\frac{3a_{2}}{2}\|B_{R}\| \leq R$  we obtain:
$$
\|f_{A_{R},B_{R}}(\lambda)\| \leq 2 \lambda^{5} e^{R} \mbox{ , if $\lambda \leq 1$} \;.
$$

Let  $A$ and $B$ be such that $(a_{1}-a_{2})\|A\|-\frac{3a_{2}}{2}\|B\| \leq R$.
\\

By setting $A_{R}=\frac{RA}{(a_{1}-a_{2})\|A\|-\frac{3a_{2}}{2}\|B\|}$
and  $B_{R}=\frac{RB}{(a_{1}-a_{2})\|A\|-\frac{3a_{2}}{2}\|B\|}$, we obtain\\
$$
A=\frac{(a_{1}-a_{2})\|A\|-\frac{3a_{2}}{2}\|B\|}{R}A_{R}\;,\quad
B=\frac{(a_{1}-a_{2})\|A\|-\frac{3a_{2}}{2}\|B\|}{R}B_{R}
$$
and $(a_{1}-a_{2})\|A_{R}\|-\frac{3a_{2}}{2}\|B_{R}\| = R$.\\

Then
 \begin{align*}
f_{A,B}(1)&=f_{A_{R},B_{R}}\left(\frac{(a_{1}-a_{2})\|A\|-\frac{3a_{2}}{2}\|B\|}{R}\right)\\
\|f_{A,B}(1)\| &\leq 2
e^{R} \left (\frac{(a_{1}-a_{2})\|A\|-\frac{3a_{2}}{2}\|B\|}{R} \right ) ^{5}\\
&\leq 2 \frac{e^{R}}{R^{5}}\left
  ((a_{1}-a_{2})\|A\|-\frac{3a_{2}}{2}\|B\|\right)^{5} \;.\\
\end{align*}
\fin


\subsection{Error estimate of the approximated composition method}

The composition method is approximated by replacing
$e^{-i\frac{\Delta t}{\varepsilon}
  d\Gamma(-\Delta_{K})}$ by its $4^{th}$ order Taylor expansion, with
 some normalization factor.\\

Errors estimates for this modified composition method rely on the two following lemmas.

\begin{lem} \label{lem aprox split}
Let $E$ be a normed vector space, $J \in \mathbb N^*$, $(f_j)_j$ and $(g_j)_j$  two
maps sequences from $E$ to $E$ such that for all $j \in \{1,\ldots,J\}$:

\begin{itemize}

\item
$f_j$  is linear.

\item
$\|f_j(u)\|=\|g_j(u)\|=\|u\|$ for all $u \in E $.

\item
$\forall u \in E \qquad \|u\| \leq \varrho \Rightarrow
\|f_j(u)-g_j(u)\|\leq \delta$
\end{itemize}
For $u_0 \in E$, $\|u_0\| \leq \varrho$ set $u_j=f_j(u_{j-1})$ and
$v_j=g_j(v_{j-1})$ with $v_0=u_0$.\\
Then we deduce $\|u_J-v_J\| \leq J \delta$ .
\end{lem}

\proof
Let us proceed by  induction on $J$.\\
For $J=0$, $\|u_0-v_0\|=0\leq 0 \delta$.\\
Let us assume $\|u_J-v_J\| \leq J \delta$ with the hypotheses
fullfilled for $j \in \{1,\ldots,J+1\}$.\\

\begin{align*}
u_{J+1}-v_{J+1}&=f_{J+1}(u_J)-g_{J+1}(v_J)\\
&=f_{J+1}(u_J)-f_{J+1}(v_J)+f_{J+1}(v_J)-g_{J+1}(v_J) \;.\\
\end{align*}

Since  $f_{J+1}$ is linear and unitary we obtain:

$$
\|f_{J+1}(u_J)-f_{J+1}(v_J)\|=\|f_{J+1}(u_J-v_J)\|=\|u_J-v_J\|\leq
J \delta \;.
 $$

Moreover $\|v_J\|=\|g_J \circ g_{J-1} \circ \ldots \circ
g_1(u_0)\|=\|u_0\| \leq \varrho$ ,\\

then
$$
\|f_{J+1}(v_J)-g_{J+1}(v_J)\|\leq \delta \;,
$$

we obtain
$$
\|u_{J+1}-v_{J+1}\|\leq (J+1)\delta \;.
$$
\fin
\\

Let $TL(e^{A})$ denote the $4^{th}$ order Taylor expansion of $e^{A}$
around $0$.\\
Let $A$ be in $B(0,c_{R})$, $c_{R} >0$.\\
 
\begin{lem} \label{TLtilde}
 Let $u$ be a vector in a normed vector space $E$ and let $A$ be an anti-adjoint
 operator on $E$. Define the application
$\widetilde{TL}(e^{A})$ on $E$ which is non linear by:
$$
\widetilde{TL}(e^{A})u=\frac{\|u\|}{\|TL(e^{A})u\|} TL(e^{A})u  \mbox{    if
  $\|TL(e^{A})u\| \neq 0$, }
$$
it preserves the norm.\\
Then
$\|(TL(e^{A})-\widetilde{TL}(e^{A}))u\| \leq \|TL(e^{A})-e^{A}\| \|u\|$.
\end{lem}

\proof

\begin{align*}
TL(e^{A})u-\widetilde{TL}(e^{A})u&=TL(e^{A})u-\frac{\|u\|}{\|TL(e^{A})u\|}TL(e^{A})u\\
&=\left (1-\frac{\|u\|}{\|TL(e^{A})u\|} \right ) TL(e^{A})u
\end{align*}

\begin{align*}
\|TL(e^{A})u-\widetilde{TL}(e^{A})u\|&=\Big\| \left
  (1-\frac{\|u\|}{\|TL(e^{A})u\|} \right ) TL(e^{A})u \Big\|\\
&=|\|TL(e^{A})u\|-\|u\||\\
&=|\|TL(e^{A})u\|-\|e^{A} u\||\\
&\leq \|TL(e^{A})u-e^{A} u\|\\
\|(TL(e^{A})-\widetilde{TL}(e^{A}))u\| &\leq \|TL(e^{A})-e^{A}\| \|u\| .
\end{align*}

\fin
\\

The $4^{th}$ order error of the Taylor expansion gives:
\begin{align*}
\|e^{A}-TL(e^{A})\|= \left \|\sum_{i=5}^{\infty}\frac{A^{i}}{i!}
\right\| \leq \|A\|^{5}\int_{0}^{1}\frac{(1-t)^{4}}{4!}\|e^{tA}\|dt \leq \|A\|^{5}\frac{1}{5!}\,.
\end{align*}

The following proposition gives an estimate of the approximation of
the composition method.\\

\begin{prop}
Let $A$ and $B$ be two anti-adjoint matrices and $J$ an integer such that
$$
\frac{\Delta t}{\varepsilon} ((a_{1}-a_{2})\|A\|-\frac{3a_{2}}{2}\|B\|) \leq 5
$$
and
$$
J \geq \frac{t}{5\varepsilon}
((a_{1}-a_{2})\|A\|-\frac{3a_{2}}{2}\|B\|) \;.
$$

Then
\begin{align*}
\| e^{\frac{ t}{\varepsilon}(A+B)}u - (\widetilde \Psi_{\frac{ \Delta t}{\varepsilon} A,\frac{ \Delta t}{\varepsilon} B})^{J}u \|
 &\leq\left (2\Big(\frac{e}{5}\Big)^{5}\left ((a_{1}-a_{2})\|A\|-\frac{3a_{2}}{2}\|B\|\right)^{5}+\frac{3}{4}\|A\|^{5} \right )
 t\frac{\Delta t^{4}}{\varepsilon^{5}} \|u\| 
\end{align*}
for all vector $u$, where
$$\widetilde \Psi_{ A,
  B}=e^{\frac{a_{1}B}{2}}\widetilde{TL}(e^{a_{1}A})
e^{\frac{a_{1}B}{2}}e^{\frac{a_{2}B}{2}}\widetilde{TL}(e^{a_{2}A})
e^{\frac{a_{2}B}{2}}e^{\frac{a_{1}B}{2}}\widetilde{TL}(e^{a_{1}A})
e^{\frac{a_{1}B}{2}} \;.$$
\end{prop}

\proof

 Let $u$ be a normed vector.\\

First for $i=1,2,3$ let us estimate the error:\\
$$
\|e^{\frac{a_{i}B}{2}}e^{a_{i}A}e^{\frac{a_{i}B}{2}}u-e^{\frac{a_{i}B}{2}}\widetilde{TL}(e^{a_{i}A}) e^{\frac{a_{i}B}{2}}u\|
$$

by using the fact $e^{\frac{a_{i}B}{2}}$ is an unitary operator and Lemma \ref{TLtilde}:
\begin{align*}
\|e^{\frac{a_{i}B}{2}}e^{a_{i}A}e^{\frac{a_{i}B}{2}}u-e^{\frac{a_{i}B}{2}}\widetilde{TL}(e^{a_{i}A}) e^{\frac{a_{i}B}{2}}u\|&=\|(e^{a_{i}A}-\widetilde{TL}(e^{a_{i}A}))e^{\frac{a_{i}B}{2}}u\|\\
&\leq \frac{\|a_{i}A\|^{5}}{60}  \leq \frac{\|A\|^{5}}{4} \;.\\
\end{align*}
like in the previous proof.\\

Now Lemma \ref{lem aprox split} can be applied with
$f_i=e^{\frac{a_i}{2}B}e^{a_iA}e^{\frac{a_i}{2}B}$,
$g_{i}=e^{\frac{a_{i}B}{2}}\widetilde{TL}(e^{a_{i}A}) e^{\frac{a_{i}B}{2}}$
and $J=3$.\\

Then 
$$
\|\Psi_{A,B}u-\widetilde\Psi_{A,B}u \| \leq \frac{3}{4}\|A\|^{5} \;.
$$

Secondly let us estimate the error:\\
$$
\| e^{A+B}u -\widetilde\Psi_{A,B}u \|
$$\\

 By using Proposition \ref{comp estimates} with its hypotheses and the previous estimates:
\begin{align*}
\| e^{A+B}u -\widetilde\Psi_{A,B}u\|
&\leq
\|e^{A+B}-\Psi_{A,B}\|
\|u\|+\|(\Psi_{A,B}-\widetilde \Psi_{A,B})u\|\\
& \leq \frac{2e^{R}}{R^{5}}\left
  ((a_{1}-a_{2})\|A\|-\frac{3a_{2}}{2}\|B\|\right)^{5}+\frac{3}{4}\|A\|^{5}
\;. \\
\end{align*}

By applying that to $\frac{\Delta t}{\varepsilon}A$ and $\frac{\Delta
  t}{\varepsilon}B$ where $\Delta t=\frac{t}{J}$ with $J$ positive integer, we obtain:

\begin{align*}
\| e^{\frac{\Delta t}{\varepsilon}(A+B)}u-\widetilde \Psi_{\frac{ \Delta t}{\varepsilon} A,\frac{ \Delta t}{\varepsilon} B}u \|
 & \leq \left (\frac{2e^{R}}{R^{5}}\left ((a_{1}-a_{2})\|A\|-\frac{3a_{2}}{2}\|B\|\right)^{5}+\frac{3}{4}\|A\|^{5} \right ) \frac{\Delta
t^{5}}{\varepsilon^{5}} \;.\\
\end{align*}

Then by applying Lemma \ref{lem aprox split} with
$f_{j}=e^{\frac{\Delta t}{\varepsilon}(A+B)}$ and $g_{j}=\widetilde \Psi_{\frac{ \Delta t}{\varepsilon} A,\frac{ \Delta t}{\varepsilon} B}$, we obtain:

\begin{align*}
\| e^{\frac{ t}{\varepsilon}(A+B)}u - (\widetilde \Psi_{\frac{ \Delta t}{\varepsilon} A,\frac{ \Delta t}{\varepsilon} B})^{J}u \|
 &\leq \left (\frac{2e^{R}}{R^{5}}\left ((a_{1}-a_{2})\|A\|-\frac{3a_{2}}{2}\|B\|\right)^{5}+\frac{3}{4}\|A\|^{5} \right )
 t\frac{\Delta t^{4}}{\varepsilon^{5}} \;.\\
\end{align*}




 
By knowing that for all positive integer $\tau$ and positive real $a$, 
$R \mapsto \frac{e^{aR}}{R^{\tau}}$  is minimal in   $R_{min}=\frac{\tau}{a}$
and $\min_{R>0}\frac{e^{aR}}{R^{\tau}}=(\frac{ae}{\tau})^{\tau}$ , the condition

$$
\frac{\Delta t}{\varepsilon} \left
  ((a_{1}-a_{2})\|A\|-\frac{3a_{2}}{2}\|B\|\right) \leq 5 \;,
$$
that is
$$
J \geq \frac{t}{5\varepsilon}
((a_{1}-a_{2})\|A\|-\frac{3a_{2}}{2}\|B\|) \;,
$$
implies
\begin{align*}
\| e^{\frac{ t}{\varepsilon}(A+B)}u - (\widetilde \Psi_{\frac{ \Delta t}{\varepsilon} A,\frac{ \Delta t}{\varepsilon} B})^{J}u \|
 &\leq \left (2\left(\frac{e}{5}\right )^{5}\left ((a_{1}-a_{2})\|A\|-\frac{3a_{2}}{2}\|B\|\right)^{5}+\frac{3}{4}\|A\|^{5} \right )
 t\frac{\Delta t^{4}}{\varepsilon^{5}} \;.\\
\end{align*}
\fin

For $\widetilde Q\in \mathcal L (\bigvee^2 \mathcal Z)$, we know according
to \eqref{op.wick}:
\begin{align*}
Q^{Wick}_{|\bigvee^n \mathcal Z}&=\frac{\sqrt{n!(n+2-2)!}}{(n-2)!}\varepsilon^{\frac{2+2}{2}}S_{n-2+2}(\widetilde
Q \otimes Id^{\otimes n-2})\\
&=\varepsilon^2n(n-1)S_n(\widetilde Q \otimes Id^{\otimes n-2}) \;,
\end{align*}
then
\[ \|Q^{Wick}_{|\bigvee^N \mathcal Z}\| \leq \varepsilon^2N(N-1) \|S_N\| \|\widetilde Q\| \leq
\varepsilon^2N^2\|\widetilde Q\| = \|\widetilde Q\| \;. \]
When $Q^{Wick}=\mathcal V$ with
$$
\widetilde{\mathcal V} (e_i \vee e_j)=\frac12 V_{ij} e_i \vee e_j \;,
$$ 

the norm $\|\mathcal V\|$ is bounded from above by $\|\widetilde{\mathcal V}\|=\frac12
\max |V_{ij}|$ independently of the number $N=\lfloor \frac
1{\varepsilon} \rfloor $ of particles.\\

Moreover
\( \|d\Gamma(A)_{|\bigvee^N \mathcal Z}\| \leq \varepsilon N
\|A\|=\|A\| \;, \) therefore $\|d\Gamma(-\Delta_K)\| \leq \|\Delta_K\|=2$.
\\
Finally by applying the last proposition with $A=-\mathrm i
d\Gamma(-\Delta_K)$ and $B=-\mathrm i \mathcal V$, an error estimate is obtained for the complete evolution:
\begin{align*}
\| e^{-\frac{\mathrm i t}{\varepsilon}H_{\varepsilon}}u - (\widetilde \Psi_{-\frac{ \Delta t}{\varepsilon}\mathrm i
d\Gamma(-\Delta_K),-\frac{ \Delta t}{\varepsilon} \mathrm i \mathcal V})^{J}u \|
 &\leq \left (2\left(\frac{e}5\right)^{5}\left ((a_{1}-a_{2})\|\Delta_K\|-\frac{3a_{2}}{2}\|\widetilde{\mathcal V}\|\right)^{5}+\frac{3}{4}\|\Delta_K\|^{5} \right )
 t\frac{\Delta t^{4}}{\varepsilon^{5}} \;.\\
\end{align*}

Pratically, the time step is chosen according to $N$ and $t$ so that the above error is negligeable.

\section{Numerical simulations}
\label{sec.num-simu}

For all the numerical simulations the final time is chosen to be
$t_{max}=1$, the number of time steps for the $4^{th}$ order
Runge-Kutta method applied to solve the mean field equation is $100$.\\
The loop of the number of particles is performed numerically from
$N_{min}=2$ 
to $N_{max}=20$ particles, and only for an even number of particles.\\
In the Fortran program the computations were performed by parallelizing the loop in the
computation of the product sparse matrix-vector $d\Gamma(-\Delta_K)u$
with Openmp on 8 threads.\\

{\bf Results and orders of convergence for $\gamma_{\varepsilon}^{(1)}$ and $\gamma_{\varepsilon}^{(2)}$}
\\

For each type of states, the following graphics show for the reduced density matrices and for $K=10$ sites:

\begin{itemize}
\item[1)]
The logarithm of the error in trace norm $\log(\max_{t\in[0,1]}\left\|\gamma^{(p)}_{N}(t)-\gamma_{\infty}^{(p)}(t)\right\|_1)$
according to the logarithm of the number of particles $N$ in the cases
$p=1$ and $2$.\\
A straight line is obtained whose the slope is the order of the error in
$1/N$.\\
These numerical experiments also valid the idea that for
rather smooth but non trivial $N$-body bosonic system, the mean field
asymptotics start to be relevant at $N=4$. The numerical plot agree perfectly with the theoretical results of
\cite{AFP}. 
\item[2)]
In the case $p=1$ the density of particles on each site $k \in\{1,\ldots,K\}$ given by
$\gamma^{(1)}_{kk}(t)$ for $N=20$ particles and for the mean field limit at the same times $t=0$ and $t=1$.\\
\item[3)]
The correlations in terms of the $1$ and $2$ particles reduced density
matrices, for $N=20$ particles and the mean field at the time
$t=1$. Depending on the case, this plot shows with which accuracy the
mean field also catches some quantum correlations.
\end{itemize}

\subsection{Hermite states}

For the Hermite states $z^{\otimes N}$ the vector $z$ is given by 
$z=\frac1{\sqrt{3}}((1+\mathrm i)e_1+\mathrm i e_3)$.\\

\begin{figure}[H]
  \centering
\subfloat[][Log-log plot for Hermite states. p=1, K=10.\\Numerical slope -0,9804]
{\label{fig:log-log-Hermite}
\includegraphics[scale=0.45]{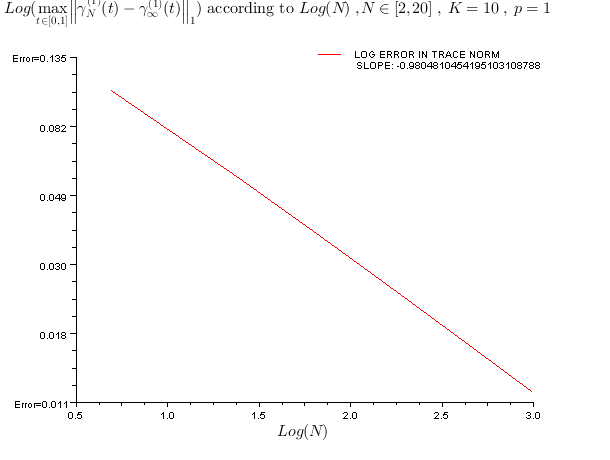}}
\subfloat[][Compared densities of particles at times t=0 and t=1]
{\label{fig:dens-part-Hermite}
\includegraphics[scale=0.45,width=8cm]{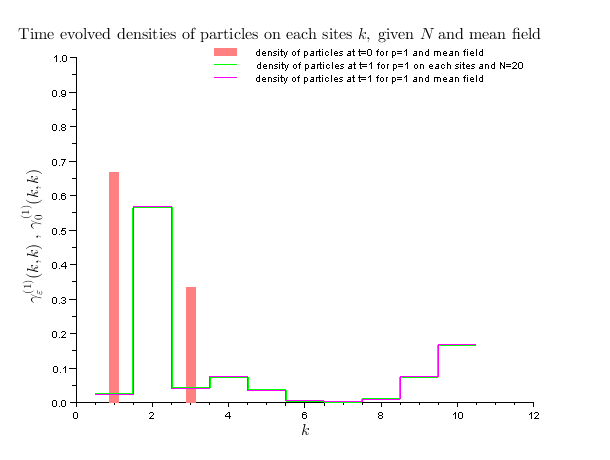}}
\end{figure}

\begin{figure}[H]
  \centering
\subfloat[][Log-log plot for Hermite states. p=2, K=10.\\Numerical slope -0,9693]
{\label{fig:log-log-Hermitestates-p2}
\includegraphics[scale=0.45]{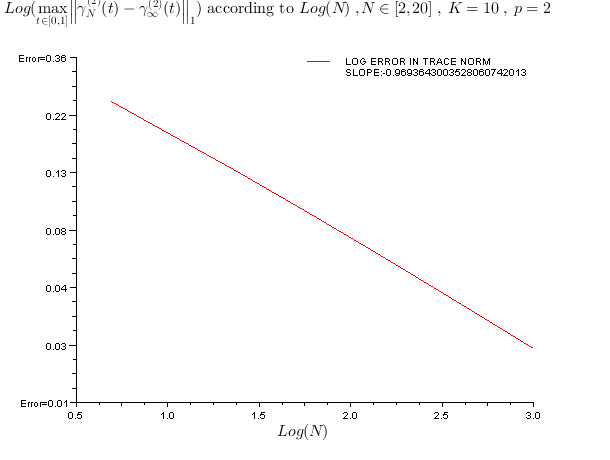}}
\subfloat[Mean field(white) and 20-body quantum(blue) correlations at time $t=1$]
{\label{fig:corr-dens-part-Hermitestates-p2p1}
\includegraphics[scale=0.45,width=8cm]{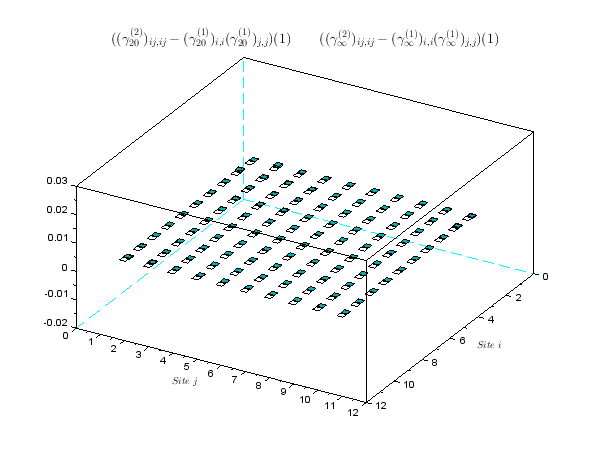}}
\end{figure}

\subsection{Twin states}

For the twin states 
$\Psi_N=\frac{a^*(\psi_1)^{n_1}a^*(\psi_2)^{n_2}}{\sqrt{\varepsilon^{n_1+n_2}n_1!n_2!}}|\Omega\rangle$,
$\psi_1=\frac1{\sqrt{2}}(e_1+\mathrm i e_3)$ and $\psi_2=e_2$.\\

\begin{figure}[H]
  \centering
  \subfloat[][Log-log plot for twin states. p=1, K=10.\\ Numerical slope -0,9855]{\label{fig:log-log-twinstate}\includegraphics[scale=0.45]{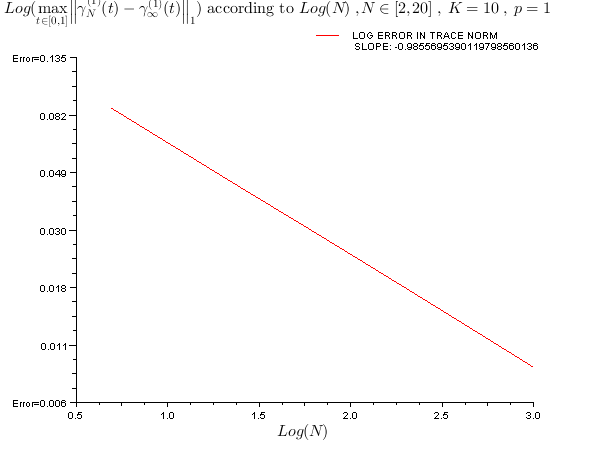}}
  \subfloat[Compared densities of particles at times t=0 and t=1]{\label{fig:dens-part-twinstate}\includegraphics[scale=0.45,width=8cm]{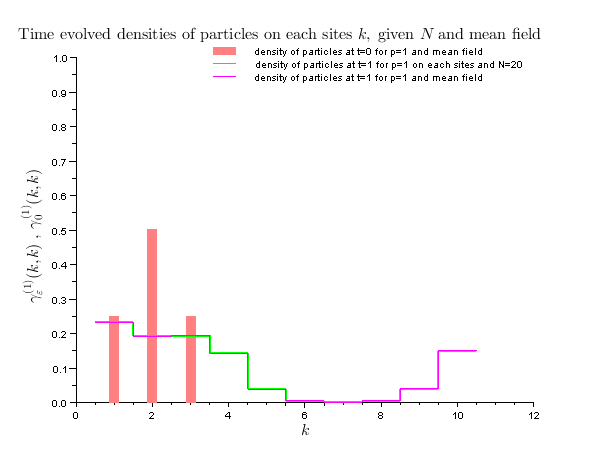}}
\end{figure}

\begin{figure}[H]
  \centering
  \subfloat[][Log-log plot for twin states. p=2, K=10.\\ Numerical slope -1,0566]{\label{fig:log-log-twinstates-p2}\includegraphics[scale=0.45]{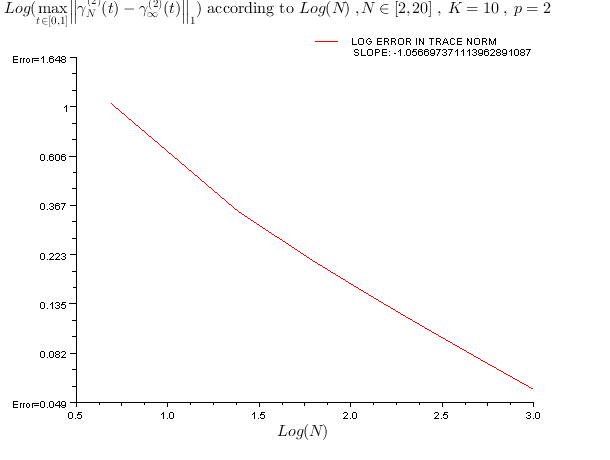}}
\subfloat[Mean field(white) and 20-body quantum(blue) correlations at time $t=1$]{\label{fig:corr-dens-part-twin states-p2p1}\includegraphics[scale=0.45,width=8cm]{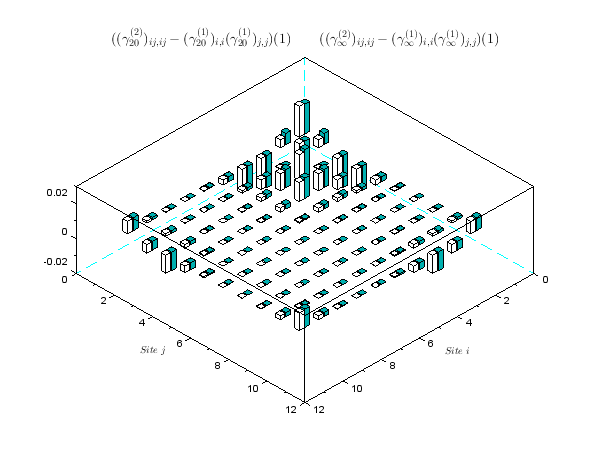}}
\end{figure}

\subsection{Wq states}

For the Wq states 
$\Psi_N=\frac{a^*(\psi_1)^{n_1}a^*(\psi_2)^{n_2}}{\sqrt{\varepsilon^{n_1+n_2}n_1!n_2!}}|\Omega\rangle$,
$\psi_1=\frac1{\sqrt{2}}(e_1+\mathrm i e_3)$ and $\psi_2=e_2$.\\

In this case the state is given by $S_N(\psi_1^{\otimes n_1}\otimes
\psi_2^{\otimes n_2})$, with $n_1+n_2=N$, $n_1=N-q$ and $n_2=q$ fixed
for the mean field.\\
The associated Wigner measure is $\delta^{S^1}_{\psi_1}$.\\
In these simulations $q=2$.

\begin{figure}[H]
  \centering
  \subfloat[][Log-log plot for Wq states. p=1, K=10.\\ Numerical slope -0,9843]{\label{fig:log-log-Wq}\includegraphics[scale=0.45]{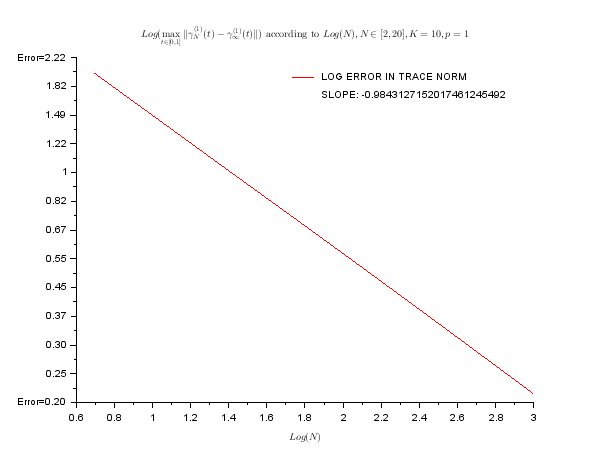}}
  \subfloat[Compared densities of particles at times t=0 and t=1]{\label{fig:dens-part-Wq}\includegraphics[scale=0.45,width=8cm]{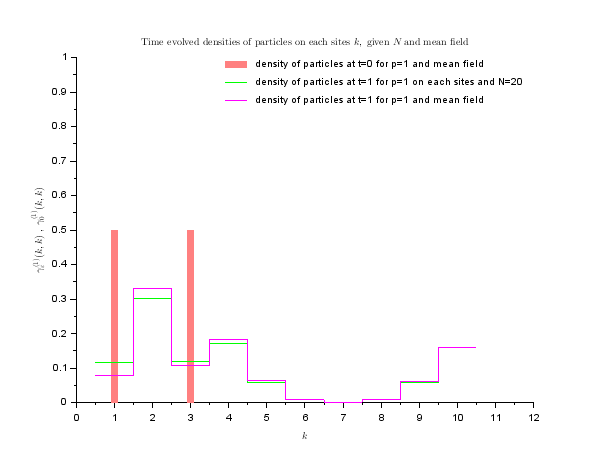}}
\end{figure}

\begin{figure}[H]
  \centering
  \subfloat[][Log-log plot for Wq states. p=2, K=10.\\ Numerical slope
  -0,9449]{\label{fig:log-log-Wqstates-p2}\includegraphics[scale=0.45]{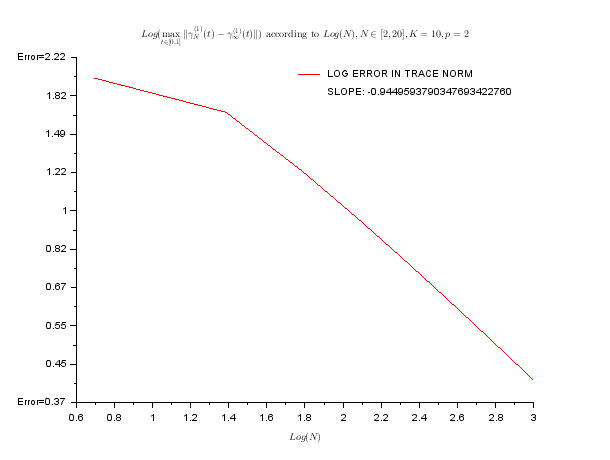}}
  \subfloat[Mean field(white) and 20-body quantum(blue) correlations at time $t=1$]{\label{fig:corr-dens-part-Wq-p2p1}\includegraphics[scale=0.45,width=8cm]{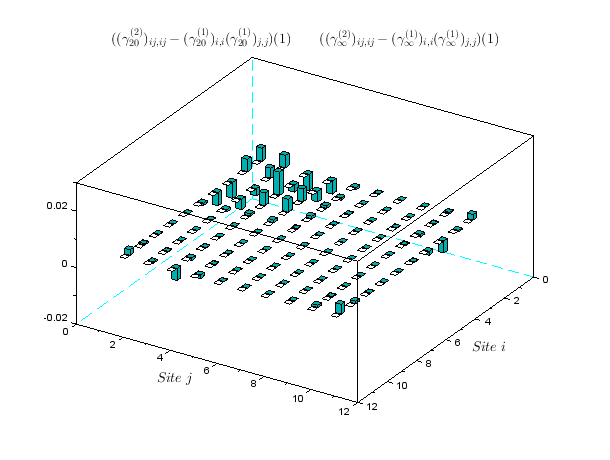}}
\end{figure}

\subsection{Other states}

A case when the order of convergence is equal to $1/2$.(see\cite{AFP})\\
In this case $\varrho_{\varepsilon}=|\phi_N^{\otimes N} \rangle
\langle \phi_N^{\otimes N}|$ with
$\phi_N=\frac1{\sqrt{N}}e_1+\sqrt{1-\frac1{N}}e_2$.\\
The associated Wigner measure is $\delta^{S^1}_{e_2}$. 

\begin{figure}[H]
  \centering
  \subfloat[Log-log plot. p=1, K=10. Numerical slope -0,4711]{\label{fig:log-log-states}\includegraphics[scale=0.45]{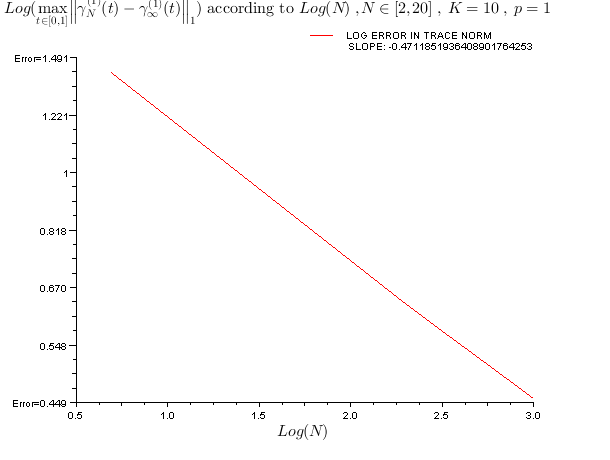}}
  \subfloat[Compared densities of particles at times t=0 and t=1]{\label{fig:dens-part-states}\includegraphics[scale=0.45,width=8cm]{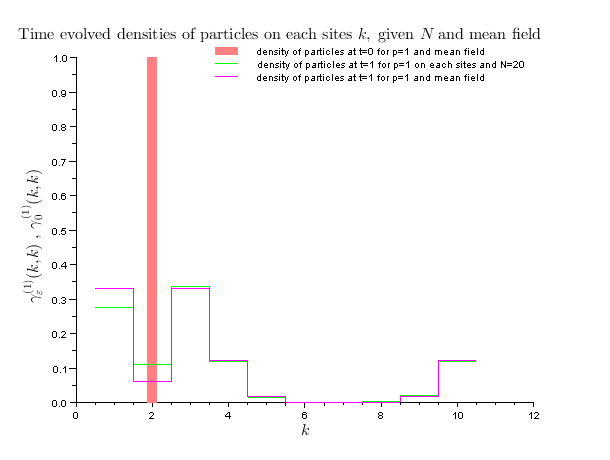}}
\end{figure}

\begin{figure}[H]
  \centering
  \subfloat[Log-log plot. p=2, K=10. Numerical slope -0,4524]{\label{fig:log-log-zn0.5states-p2}\includegraphics[scale=0.45]{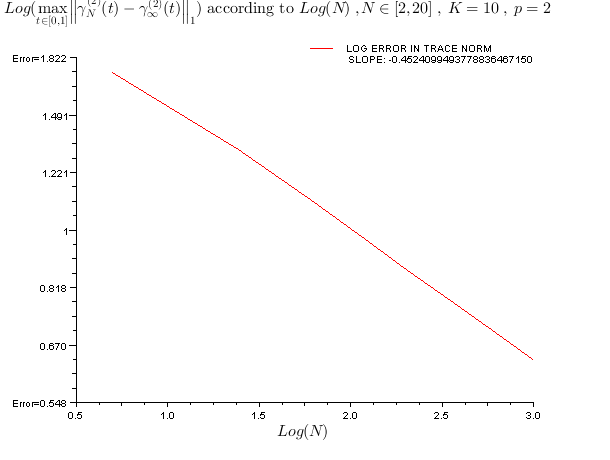}}
  \subfloat[Mean field(white) and 20-body quantum(blue) correlations at time $t=1$]{\label{fig:corr-dens-part-twin states-p2p1}\includegraphics[scale=0.45,width=8cm]{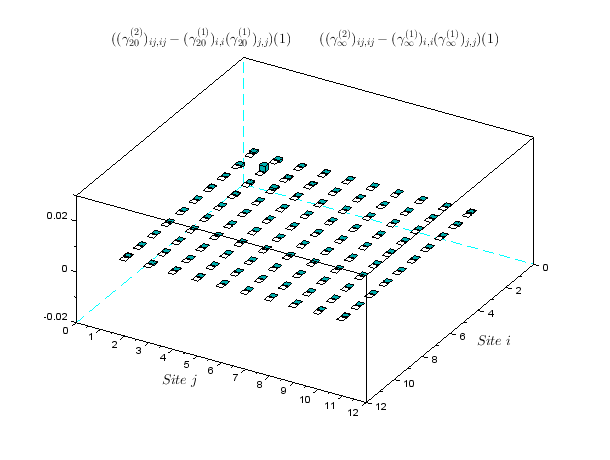}}
\end{figure}

\newpage
\section*{Appendix}
\label{sec.appendix}
{\bf Class of symbols} \cite{AmNi1,AmNi2,AmNi3,AmNi4}.
For any $p,q\in \mathbb{N}$, define $\P_{p,q}$ to be the space
 of homogeneous complex-valued polynomials on $\Z$  such that $b\in\P_{p,q}$
 if and only if there exists a (unique) bounded operator $\tilde b\in\L(\vee^p\Z,\vee^q\Z)$ such that for all $z\in\Z$:
\bea
\label{eq.wick}
b(z)=\la z^{\otimes q}, \tilde{b} \,z^{\otimes p}\ra\,.
\eea  
\begin{align}\label{op.wick}
b^{Wick} \,_{|\vee^n \Z}=1_{[p,+\infty)}(n)\frac{\sqrt{n!
(n+q-p)!}}{(n-p)!} \;\hbarr^{\frac{p+q}{2}} \;\S_{n-p+q}\left(\tilde{b}\otimes 1^{\otimes(n-p)}\right)\,,
\end{align}
where $\tilde{b}$ denotes the operator associated with the symbol $b$ according to
\eqref{eq.wick}.

 The composition method based on the Strang splitting with the
coefficients \eqref{coeff.composition} is of $4^{th}$ order (see \cite{HLW}).
\\

Dimension of $\bigvee^N \mathbb C^K$: $\dbinom{N+K-1}{K-1}$ for $K$ in
$[1,10]$ and $N$ in $[1,20]$:\\

\label{dimensions}
\begin{tabular}{||l||*{9}{c|}l|} 
  \hline
  & $K=1$ & 2&3&4&5&6&7&8&9&10 \\
  \hline 
  $N=1$ &    1&    2&     3&      4&       5&        6& 7&         8&         9&          10 \\ 
  \hline 
  2&     1&    3&     6&      10 &     15&       21& 28&  36 &       45&        55       \\ 
  \hline 
  3&     1&    4&     10 &    20 &     35 &      56  &     84&120 &      165&        220 \\ 
  \hline   
  4&     1&    5&     15 &    35 &     70 &      126 &     210&330 &      495&        715   \\ 
  \hline   
  5&     1&    6&     21 &    56  &    126 &     252 &     462& 792 &      1287 &      2002 \\ 
  \hline  
  6&     1&    7&     28 &    84 &     210 &     462 &     924& 1716 &     3003&       5005   \\ 
  \hline 
  7 &    1&    8&     36 &    120 &    330 &     792 &     1716& 3432 &     6435 &      11440  \\ 
  \hline 
  8  &   1 &   9 &    45&     165 &    495 &     1287 &    3003& 6435&      12870  &    24310 \\ 
  \hline  
  9&     1&    10 &   55&     220 &    715 &     2002 &    5005&    11440&     24310&      48620   \\ 
  \hline 
  10 &   1 &   11&    66 &    286 &    1001&     3003  &   8008&  19448 &    43758  &    92378   \\ 
  \hline 
  11&    1&    12 &   78&     364&     1365&     4368&     12376&31824&     75582 &     167960 \\ 
  \hline 
  12&    1&    13&    91&     455&     1820&     6188 &    18564& 50388 &    125970&     293930  \\ 
  \hline  
  13&    1&    14&    105&    560&     2380&     8568&     27132&77520&     203490&     497420  \\ 
  \hline  
  14&    1&    15 &   120&    680&     3060&     11628&    38760&116280   & 319770   &  817190 \\ 
  \hline   
  15 &   1&   16&    136 &   816&     3876&     15504&    54264&170544   & 490314    & 1307504 \\ 
  \hline  
  16 &   1 &   17  &  153   & 969     &4845  &   20349 &   74613& 245157&    735471 &    2042975 \\ 
  \hline 
  17  &  1  &  18  &  171&    1140  &  5985&     26334   & 100947& 346104    &1081575&   3124550 \\ 
  \hline  
  18   & 1    &19&    190&    1330&    7315  &   33649  &  134596 & 480700   & 1562275&    4686825 \\ 
  \hline   
  19   & 1  &  20  &  210   & 1540 &   8855     & 42504 &   177100 &657800   & 2220075   & 6906900 \\  
  \hline  
  20   & 1 &   21 &   231&   1771&    10626&    53130&    230230&888030  &  3108105 &   10015005 \\
  \hline 
\end{tabular} 

\bigskip
\noindent
\medskip

\newpage
\section*{Acknowledgement}
The research has been supported by the ANR-11-IS01-0003
Lodiquas.\\
The numerical computations were performed on the servers VSC of the
University of Vienna and MAGI of the University Paris 13. The numerical illustrations were obtained with Scilab.\\
I thank Gilles Vilmart, Gilles Scarella and Hans Peter Stimming for
useful discussions about numerical schemes and computers. I also thank my advisors Francis Nier and Norbert
J. Mauser for their remarks and commitment for the writing of this article.

\bibliographystyle{empty}

\end{document}